\documentclass[times,twocolumn,final,longtitle]{elsarticle}

\usepackage{jasr}
\usepackage{framed,multirow}
\usepackage{threeparttable}

\usepackage{amssymb}
\usepackage{amsmath}
\usepackage{bm}
\usepackage{latexsym}
\usepackage{xspace}

\usepackage{lineno}

\usepackage{xcolor}
\definecolor{newcolor}{rgb}{1.0,.349,.1}
\definecolor{red}{rgb}{0.8,0.0,0.50}

\usepackage[colorlinks,linkcolor=magenta,citecolor=blue,filecolor=cyan,urlcolor=purple]{hyperref}

\graphicspath{{./}{figures/}}

\journal{Advances in Space Research}

\begin{document}

\verso{C.~Verbeke \textit{et al.}}

\begin{frontmatter}


\title{Quantifying errors in 3D CME parameters derived from synthetic data using white-light reconstruction techniques}

\author[1]{Christine~\snm{Verbeke}\corref{cor1}}
\cortext[cor1]{Corresponding author}
  \ead{christine.verbeke@oma.be}
\author[2]{M.~Leila~\snm{Mays}}
\author[2,3]{Christina~\snm{Kay}}
\author[4]{Pete~\snm{Riley}}
\author[4]{Erika~\snm{Palmerio}}
\author[5]{Mateja~\snm{Dumbovi{\'c}}}
\author[1,6]{Marilena~\snm{Mierla}}
\author[7,8]{Camilla~\snm{Scolini}}
\author[9]{Manuela~\snm{Temmer}}
\author[10,11]{Evangelos~\snm{Paouris}}
\author[2,10]{Laura~A.~\snm{Balmaceda}}
\author[12]{Hebe~\snm{Cremades}}
\author[13,9]{J{\"u}rgen~\snm{Hinterreiter}}

\address[1]{Solar--Terrestrial Centre of Excellence---SIDC, Royal Observatory of Belgium, 1180 Brussels, Belgium}
\address[2]{Heliophysics Science Division, NASA Goddard Space Flight Center, Greenbelt, MD 20771, USA}
\address[3]{Department of Physics, The Catholic University of America, Washington, DC 20064, USA}
\address[4]{Predictive Science Inc., San Diego, CA 92121, USA}
\address[5]{Hvar Observatory, Faculty of Geodesy, University of Zagreb, HR-10000 Zagreb, Croatia}
\address[6]{Institute of Geodynamics of the Romanian Academy, 020032 Bucharest-37, Romania}
\address[7]{Space Science Center, University of New Hampshire, Durham, NH 03824, USA}
\address[8]{CPAESS, University Corporation for Atmospheric Research, Boulder, CO 80301, USA}
\address[9]{Institute of Physics, University of Graz, 8010 Graz, Austria}
\address[10]{George Mason University, Fairfax, VA 22030, USA}
\address[11]{Applied Physics Laboratory, Johns Hopkins University, Laurel, MD 20723, USA}
\address[12]{Grupo de Estudios en Heliof{\'i}sica de Mendoza, CONICET, Universidad de Mendoza, 5500 Mendoza, Argentina}
\address[13]{Space Research Institute, Austrian Academy of Sciences, 8042 Graz, Austria}

\received{DD/MM/YYYY}
\finalform{DD/MM/YYYY}
\accepted{DD/MM/YYYY}
\availableonline{DD/MM/YYYY}


\begin{abstract}
Current efforts in space weather forecasting of CMEs have been focused on predicting their arrival time and magnetic structure. To make these predictions, methods have been developed to derive the true CME speed, size, position, and mass, among others.  Difficulties in determining the input parameters for CME forecasting models arise from the lack of direct measurements of the coronal magnetic fields and uncertainties in estimating the CME 3D geometric and kinematic parameters after eruption. White-light coronagraph images are usually employed by a variety of CME reconstruction techniques that assume more or less complex geometries.  This is the first study from our International Space Science Institute (ISSI) team ``Understanding Our Capabilities in Observing and Modeling Coronal Mass Ejections'', in which we explore how subjectivity affects the 3D CME parameters that are obtained from the Graduated Cylindrical Shell (GCS) reconstruction technique, which is widely used in CME research. To be able to quantify such uncertainties, the ``true'' values that are being fitted should be known, which are impossible to derive from observational data. We have designed two different synthetic scenarios where the ``true'' geometric parameters are known in order to quantify such uncertainties for the first time. We explore this by using two sets of synthetic data: 1) Using the ray-tracing option from the GCS model software itself, and 2) Using 3D magnetohydrodynamic (MHD) simulation data from the Magnetohydrodynamic Algorithm outside a Sphere (MAS) code. Our experiment includes different viewing configurations using single and multiple viewpoints. CME reconstructions using a single viewpoint had the largest errors and error ranges overall for both synthetic GCS and simulated MHD white-light data. As the number of viewpoints increased from one to two, the errors decreased by approximately 4$^\circ$ in latitude, 22$^\circ$ in longitude, 14$^\circ$ in tilt, and 10$^\circ$ in half-angle. Our results quantitatively show the critical need for at least two viewpoints to be able to reduce the uncertainty in deriving CME parameters.  We did not find a significant decrease in errors when going from two to three viewpoints for our specific hypothetical three spacecraft scenario using synthetic GCS white-light data. As we expected, considering all configurations and numbers of viewpoints, the mean absolute errors in the measured CME parameters are generally significantly higher in the case of the simulated MHD white-light data compared to those from the synthetic white-light images generated by the GCS model. We found the following CME parameter error bars as a starting point for quantifying the minimum error in CME parameters from white-light reconstructions:  
$\Delta\theta$ (latitude)=${6^\circ}_{	-	3^\circ	}^{	+	2^\circ	}$,
$\Delta\phi$ (longitude)=${11^\circ}_{	-	6^\circ	}^{	+	18^\circ	}$,
$\Delta\gamma$ (tilt)=${25^\circ}	_{	-	7^\circ	}^{	+	8^\circ	}$,
$\Delta\alpha$ (half-angle)=${10^\circ}_{	-	6^\circ}^{	+	12^\circ}$,
$\Delta h$ (height)=$0.6	_{	-	0.4	}^{	+	1.2	}$\,$R_{\odot}$,
and $\Delta\kappa$ (ratio)=$0.1	_{	-	0.02	}^{	+	0.03	}$.

\end{abstract}


\end{frontmatter}



\section{Introduction} \label{sec:intro}

Coronal mass ejections (CMEs) are large-scale eruptions of plasma and magnetic fields from the Sun that are usually regarded as the major drivers of geomagnetic effects at Earth. Current efforts in space weather forecasting of CMEs are largely focused on predicting their arrival time \citep[e.g.,][]{zhao2014,riley2018} and their magnetic structure \citep[e.g.,][]{riley2017,kilpua2019}. Other related parameters that have gained the attention of the community are CME speed, size, and mass \citep[see e.g. the review by][and references therein]{vourlidas2019}. Despite the development of numerous models and techniques aimed at predicting each of these parameters (together or separately), current predictive capabilities are still affected by relatively large errors. With regard to the current status of CME arrival time prediction, recent studies have agreed that, regardless of the method used, typical errors seem to revolve around $\pm10$~hr \citep[e.g.,][]{riley2018,wold2018,vourlidas2019}. Apart from the well-known issues related to understanding how CMEs are launched from the Sun \citep{green2018} and how they evolve through interplanetary space \citep{manchester2017,luhmann2020}, it is clear that one major drawback lies in the lack of reliable observation-based input. This includes CME input parameters (such as initial speed, geometrical properties, etc.) as well as input of the heliospheric background through which CMEs propagate, which are themselves associated with errors \citep[e.g.,][]{lee2009,hinterreiter2019}.

Difficulties in determining the input parameters for CME forecasting models arise from the lack of direct measurements of the coronal magnetic fields and uncertainties in estimating the CME geometric and kinematic parameters after eruption. When evaluating the three-dimensional (3D) CME structure and speed, it is common procedure to derive such parameters from remote-sensing data. White-light coronagraph images are usually employed by a variety of models that assume more or less complex geometries. The simplest estimates can be obtained by fitting 2D ellipses \citep[e.g.,][]{yurchyshyn2007} or 3D cones \citep[e.g.,][]{zhao2002,xie2004,xue2005} to single-viewpoint images. These techniques were initially developed to use data from the Large Angle and Spectrometric Coronagraph \citep[LASCO;][]{brueckner1995} instruments onboard the Solar and Heliospheric Observatory \citep[SOHO;][]{domingo1995}. After the launch of the Solar Terrestrial Relations Observatory \citep[STEREO;][]{kaiser2008} twin spacecraft in 2006, the coronagraphs that are part of the Sun-Earth Connection Coronal and Heliospheric Investigation \citep[SECCHI;][]{howardr2008} suite began to provide two additional viewpoints from Earth's orbital distance. Such new observations allowed the development of stereoscopic techniques to investigate the 3D structure and kinematics of CMEs. 

Triangulation methods were developed and employed, e.g., by \citet{mierla2008}, \citet{temmer2009}, \citet{liu2010}, \citet{liewer2011}, \citet{braga2017}, and \citet{balmaceda2018} to derive the de-projected kinematic properties of CMEs. Coronagraph images from two different locations can nowadays be analyzed online through the STEREO CME Analysis Tool (StereoCAT; \url{https://ccmc.gsfc.nasa.gov/analysis/stereo/}), which uses the triangulation algorithm described in \citet{mays2015}. Meanwhile, more complex geometrical models describing CME morphology emerged. For example, the Graduated Cylindrical Shell \citep[GCS;][]{thernisien2006,thernisien2009,thernisien2011} model can use both STEREO viewpoints, together with the additional view from SOHO, and an ad-hoc geometrical CME model to recover the 3D structure of CMEs in the solar corona. Three viewpoints can be employed in a variety of other models, including the Space Weather Prediction Center CME Analysis Tool \citep[SWPC\_CAT;][]{millward2013}, the empirical model of \citet{wood2009}---applied to 31 events in \citet{wood2017}, the Flux Rope in 3D \citep[FRi3D;][]{isavnin2016} model, and the revised cone model \citep{zhang2021,zhang2022}.

Several studies have focused on the comparison of different CME reconstruction methods based on coronagraph imagery. \citet{mierla2010} applied five reconstruction techniques to seven CMEs observed by the coronagraphs onboard both STEREO spacecraft in order to derive their propagation angles. They concluded that all tested methods yield latitudinal and longitudinal values that usually lie within ${\sim}10^{\circ}$ of each other, suggesting that exceptions are due to the fact that different techniques can be applied to different parts of a CME. \citet{jang2016} analyzed 306 (partial and full) halo CMEs and compared the resulting 2D parameters from single-spacecraft measurements (using plane-of-sky observations from SOHO) and 3D parameters from multi-spacecraft measurements (using StereoCAT with STEREO data). They concluded that 2D speeds are usually underestimated by ${\sim}20$\%, while 2D widths tend to be remarkably overestimated compared to their corresponding 3D parameters. Recently, \citet{paouris2021} presented a new geometrical approach for the correction of plane-of-sky CME speeds. In their study, the uncertainty in the de-projected speed estimates were bounded via upper and lower limits in the true angular width of a CME utilizing multi-viewpoint observations. The de-projected speeds were ${\sim}12.8$\% greater than their plane-of-sky speeds for the full 1037 event sample. The de-projected CME speeds were then utilized to estimate the CME arrival time at Earth also taking into account the 3D geometry of the CME. They found that slower CMEs are described better by using a spherical front to de-project the speed, whereas faster CMEs are likely flatter. However, they found only slight improvements in CME arrival time metrics when using de-projected speeds compared to using uncorrected speeds. \citet{lee2015} applied three different reconstruction methods to 44 (partial and full) halo CMEs from Earth's viewpoint, in order to establish whether the resulting 3D parameters are consistent with each other. In particular, the authors compared the single-viewpoint ice-cream cone model of \citet{xue2005}, the StereoCAT tool, and the GCS model. The study concluded that, while the two multi-viewpoint reconstructions methods are in reasonably good agreement, the cone model applied to single-spacecraft data is generally consistent with multi-spacecraft techniques when determining CME speed, but tends to underestimate the angular width for wide angular-width events (${>}100^{\circ}$). If replicated, these results could be promising, especially considering that coronagraph images from multiple viewpoints are not available at all times. Contact with the STEREO-B spacecraft was lost in late 2014 and STEREO-A data were not available during most of 2015 due to its location on the far side of the Sun. Furthermore, as STEREO-A approaches the Earth by ${\sim}22.5^{\circ}$ per year, the spacecraft has become too close (mid 2022) to provide significant additional information compared to Earth's L1 viewpoint.

In brief, it seems that different reconstruction methods yield results that are generally consistent with each other. However, it is important to note that most of the techniques mentioned above are not automated, hence the resulting parameters are highly dependent on the specific user that utilizes them. On the other hand, thus far automated or semi-automated techniques are not trustworthy enough compared to fitting by a human---e.g., automated routines may split one CME into multiple CMEs \citep{riley2018, rodriguez2022}. To the best of our knowledge, there is no published literature that explores how subjectivity affects the 3D CME parameters that are obtained from any reconstruction technique. To be able to quantify such uncertainties, one would need to know the ``true'' values that are being fitted and this is impossible to derive from observational data. To approach this problem, we have designed two different synthetic situations where the ``true'' geometric parameters are known or can be derived, in order to quantify such uncertainties for the first time. To this end, we have generated synthetic line-of-sight integrated white-light intensity images: 1) Using the ray-tracing option from the GCS model software itself, and 2) Using 3D magnetohydrodynamic (MHD) simulation data from the Magnetohydrodynamic Algorithm Outside a Sphere \citep[MAS; e.g.,][]{riley2012,riley2019} model. For the purpose of this study, we have chosen to use the GCS fitting technique, which is readily available to the community via IDL SolarSoft \citep{freeland1998}. Our analysis is performed on both single and multiple viewpoints.

It is important to consider CME reconstructions using coronagraph data within the larger context of space weather forecasting. As mentioned above, the resulting 3D parameters are often used as input for CME propagation models to estimate their arrival time and impact at Earth or other locations. While large international efforts (see e.g. the CME Arrival Time and Impact Working Team; \url{https://ccmc.gsfc.nasa.gov/assessment/topics/helio-cme-arrival.php}) are currently focused on benchmarking the performance of different propagation models \citep[e.g.,][]{verbeke2019}. However, no large studies involving many different users have been performed with the aim of quantifying the uncertainties related to CME input parameters that are usually derived from 3D reconstructions, in order to evaluate how such uncertainties affect CME propagation modeling results. In practical applications, such uncertainties will be highly valuable e.g.\ for ensemble modeling \citep{mays2015}. The work presented in this paper is a result of an International Space Science Institute (ISSI) team ``Understanding Our Capabilities in Observing and Modeling Coronal Mass Ejections'' whose goal is to explore the errors associated with CME forecasting, both from an observational point of view, such as presented within this paper, as well as for CME propagation modeling.

In this context, the work presented here aims to represent the initial step toward the creation of such benchmarking. The manuscript is organized as follows. Section~\ref{sec:synthetic} describes the methodology and results for the synthetic white-light data generated from the GCS model, while Section~\ref{sec:mhd} shows the methodology and results employed on the 3D MHD simulation data. Our results are discussed in Section~\ref{sec:discussion}, while our main conclusions are drawn in Section~\ref{sec:conclusions}.


\section{Synthetic White-light Data derived from the Graduated Cylindrical Shell model} \label{sec:synthetic}


\subsection{The Graduated Cylindrical Shell Model} \label{subsec:gcs}

The GCS model \citep{thernisien2006,thernisien2009} is an empirically-defined density model designed to reproduce the large-scale flux-rope morphology of CMEs \citep[see e.g.][for an in-depth discussion of the flux-rope morphology identifiable in coronagraph images]{vourlidas2013}. Its geometry is often referred to as a ``hollow croissant'' and consists of a half-torus frontal part with two conical legs connected back to the Sun. The resulting shape, reminiscent of a croissant, is ``hollow'' in the sense that the electron density is placed uniquely on the shell of the model. The GCS model can produce a synthetic line-of-sight integrated white-light intensity image for comparison with coronagraph images of CMEs. This means that fits performed with the GCS model can provide geometrical parameters of CMEs, but no information on their magnetic field structure. The shape and size of the croissant's shell is defined by a series of parameters that users can adjust while performing a fit: central latitude ($\theta$) and longitude ($\phi$), axial tilt ($\gamma$, calculated with respect to the solar equator), height of the apex ($h$), aspect ratio ($\kappa$, i.e., the ratio of the CME size at two orthogonal directions), and half-angle ($\alpha$, i.e., the half-angular distance between the leg axes). The model becomes equivalent to the classic ice-cream cone model \citep{fisher1984} when $\alpha$ is set to zero. Other derived parameters include the half-angle of the conical legs $\delta$, related to the aspect ratio $\kappa$ as $\kappa = \sin \delta$, as well as the face-on width $\omega_{\mathrm{FO}} = 2(\alpha + \delta)$ and edge-on width $\omega_{\mathrm{EO}} = 2\delta$, which coincide when $\alpha=0^{\circ}$ \citep{thernisien2011}. However, these widths are specific to the GCS geometry and defined as the position angle between the croissant legs which taper rapidly. Compared to the way plane-of-sky widths are measured from coronagraph CME images, the GCS widths are usually an overestimation, which becomes important when using these as inputs to CME propagation models.  For this reason we also computed the face-on width at $\beta=0^{\circ}$ \citep[see Figure 1 in][for angle definitions]{thernisien2011}. The face-on width at $\beta=0^{\circ}$ is derived from the geometry as:
\begin{equation}
\omega_{\mathrm{FO}(\beta=0^\circ)}=\mathrm{arctan}\left(\frac{\mathrm{sin}\alpha+\kappa\sqrt{1-\kappa^2+\mathrm{sin}^2\alpha}}{1-\kappa^2}\right).
\end{equation}

\citet{thernisien2006} warned about three main sources of errors in the model, the first given by errors intrinsic to the fitting method, the second being the error that stems from the photometric modeling assumptions (i.e., related to the electron density), and the third one related the results being dependent on the user’s expertise and subjective interpretation of the phenomenon. Regarding the first source of uncertainty, it is important to note that, due to the symmetry of the model, different parameters may fit the same CME well (i.e., the set of solutions may not be unique). This is especially true when fitting single-spacecraft data, which provide a single 2D projection of a 3D structure. However, when different viewpoints are available, the observations are always made from the ecliptic plane and because most solar activity comes from roughly equatorial latitudes, there are not enough constraints on the CME reconstruction. \citet{thernisien2009} performed a sensitivity analysis using a semi-automated evaluation of the GCS fit for each parameter for 26 events, finding a mean deviation of ${\pm}1.8^{\circ}$ in latitude, ${\pm}4.3^{\circ}$ in longitude, and $^{+0.07}_{-0.04}$ in aspect ratio $\kappa$. The deviations for tilt (${\pm}22^{\circ}$) and half-angle $\alpha$ ($^{+13^{\circ}}_{-7^{\circ}}$) were found to be an order of magnitude larger than those for latitude and longitude, showing that evaluation of the orientation and axial length of a CME is a particularly challenging task. This is especially true for events that are seen edge-on only (i.e., the $\alpha$ parameter is degenerate) and for those that present a distorted front and/or signatures of asymmetrical expansion, or CMEs that are seen as directly face-on only \citep[e.g.,][]{cremades2020}. 


\subsection{White-light images generated from the GCS model} \label{subsec:data-simple}

\begin{figure}[ht]
\centering
\includegraphics[width=.99\linewidth]{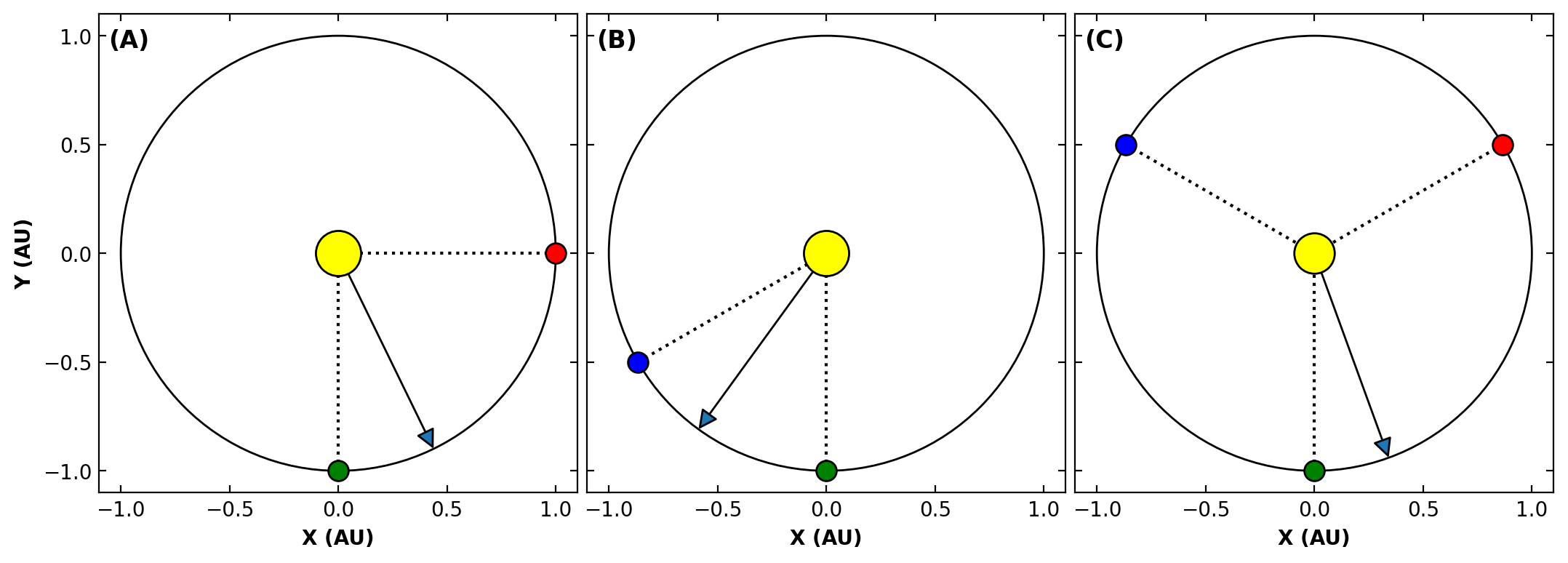}
\caption{Spacecraft configurations used for fitting the synthetic GCS-generated CMEs in Heliocentric Earth Ecliptic (HEE) coordinates. The arrow denotes the propagation direction of the CME in the ecliptic plane. A: Two spacecraft in quadrature. B: Two spacecraft separated by $60^{\circ}$, simulating two observers at the L1 and L5 points. C: Three spacecraft separated by $120^{\circ}$ from one another.}
\label{fig:config}
\end{figure}

The first set of white-light data that we study consists of images constructed via the GCS model. The GCS model is able to produce a synthetic line-of-sight integrated white-light intensity image based on a given set of the GCS model parameters. This procedure is implemented into the IDL SolarSoft routine \textit{scraytrace}, which also contains the GCS fitting routine. We have selected three different synthetic GCS CME events to fit, all of which are Earth-directed and each configuration will have a different spacecraft separation set up. In Table~\ref{tab:whitelightdetails}, we provide the GCS parameters used to create the synthetic CME events (named A, B, and C), while Figure~\ref{fig:config} shows the spacecraft separations and number of spacecraft for all three events. 

\begin{figure*}[p]
\centering
\includegraphics[width=.3\linewidth]{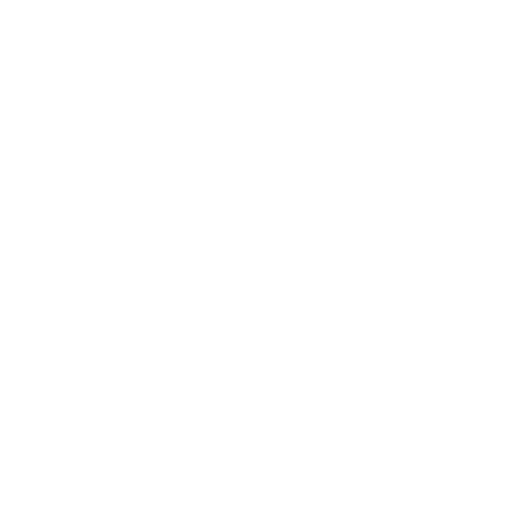}
\includegraphics[width=.3\linewidth]{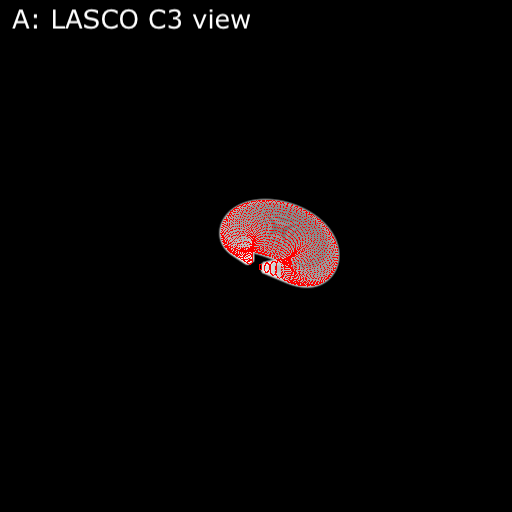}
\includegraphics[width=.3\linewidth]{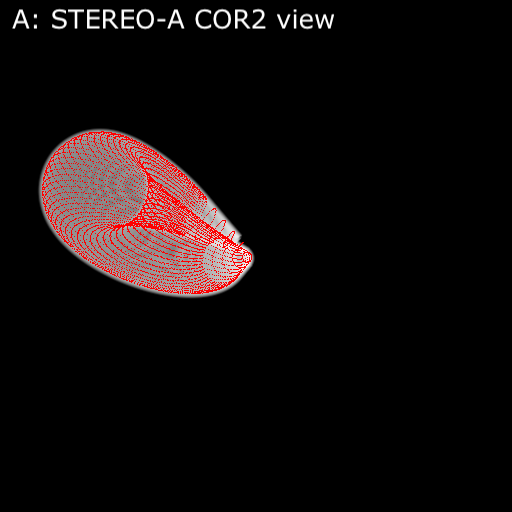}\\
\includegraphics[width=.3\linewidth]{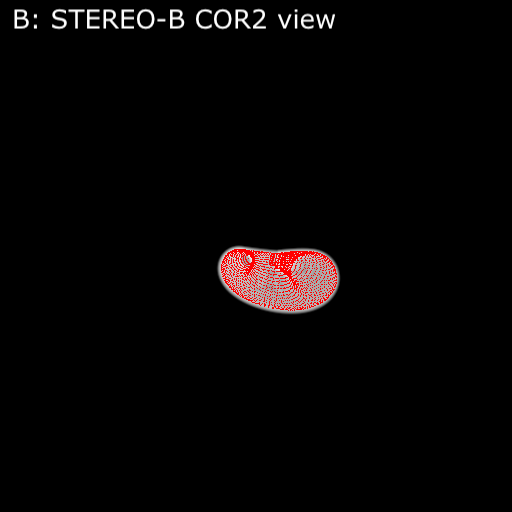}
\includegraphics[width=.3\linewidth]{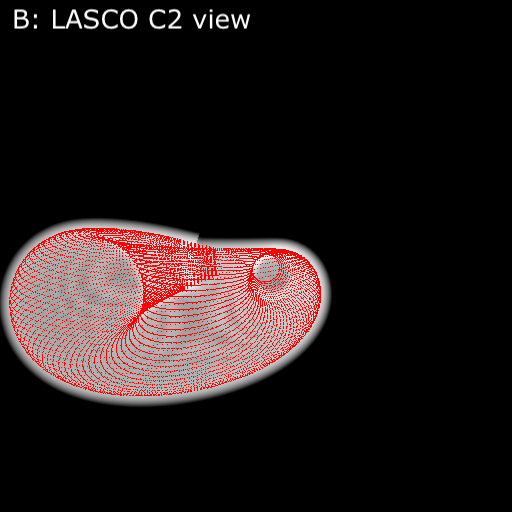}
\includegraphics[width=.3\linewidth]{figures/white_synthetic_placeholder_image.png}\\
\includegraphics[width=.3\linewidth]{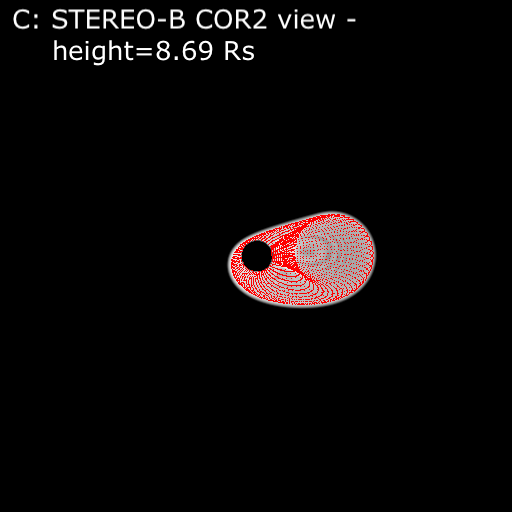}
\includegraphics[width=.3\linewidth]{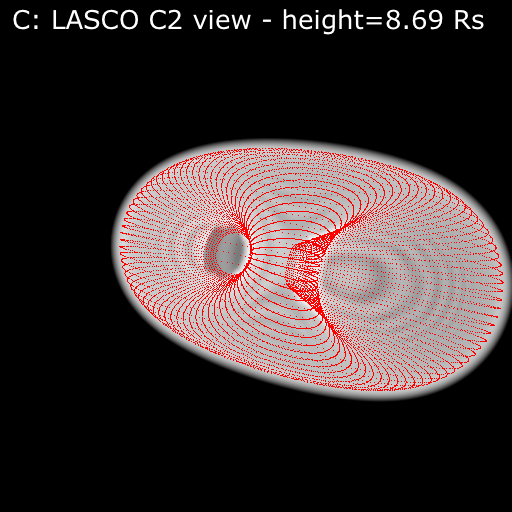}
\includegraphics[width=.3\linewidth]{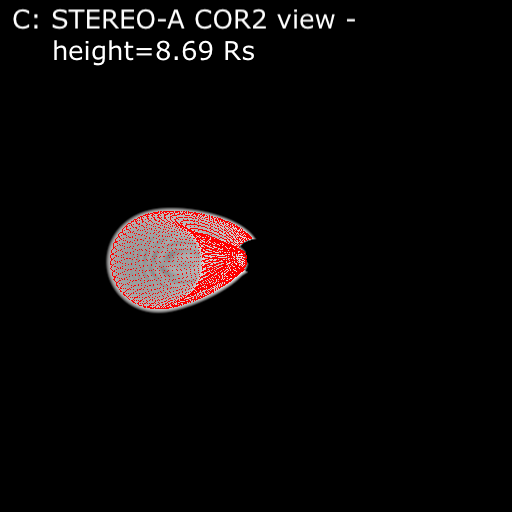}\\
\includegraphics[width=.3\linewidth]{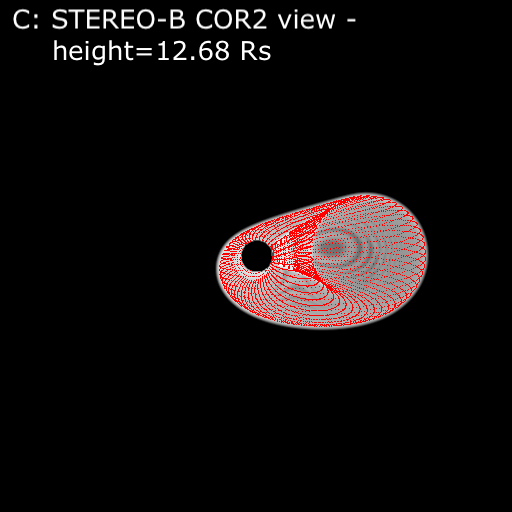}
\includegraphics[width=.3\linewidth]{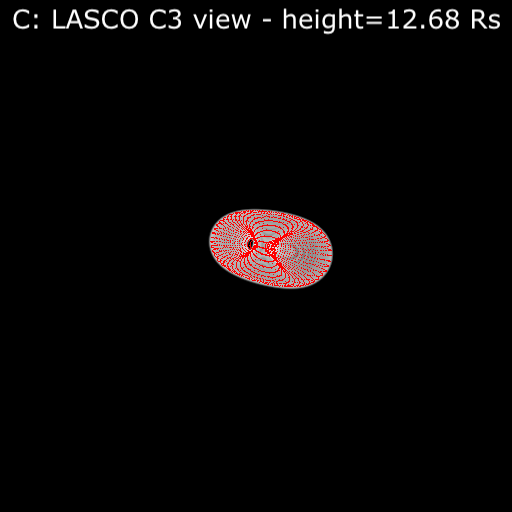}
\includegraphics[width=.3\linewidth]{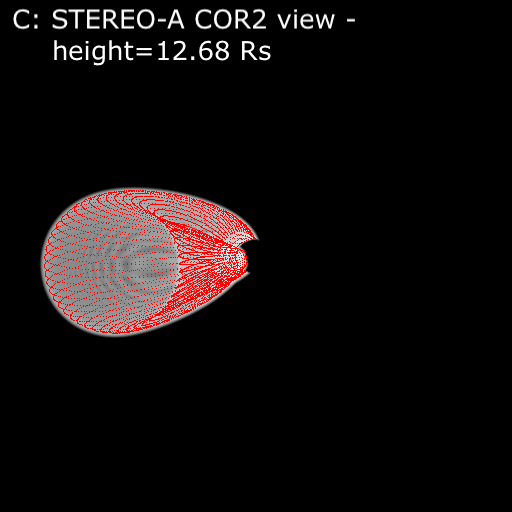}
\caption{Images of synthetic white-light data generated by GCS ray-tracing for configurations A (top row), B (second row), and C (bottom two rows), with STEREO-B (left), LASCO (middle), and STEREO-A (right) views. For configuration C, synthetic images were created for two different heights of 8.69 and 12.68 R$_{\odot}$. The true GCS fits used to create these synthetic data, are shown as the red wire frame with the parameters given in Table \ref{tab:whitelightdetails}. A schematic of the configurations is provided in Figure \ref{fig:config}.}
\label{fig:syntheticGCS}
\end{figure*}

Ten members of the team performed blind CME fits for each scenario. The reconstructions were blind in the sense that the spacecraft configurations relative to the CME direction were not revealed to the observers, and additionally the team members involved in preparing the synthetic white-light data did not perform any of the reconstructions.
For each considered CME event, we have performed a fit using only LASCO-like white-light data (i.e. the location of the spacecraft was at L1; these are scenarios A1, B1 and C1), and a fit where we also used white-light data from one to two additional spacecraft. For the first event (A), the two spacecraft are located at L1 and ${+}90^\circ$ (HEEQ) from L1 (STEREO-A location, A2 scenario); for the second (B) and third (C) events, they are located at L1 and ${-}60^\circ$ (L5, STEREO-B location, B2 scenario) and $\pm120^\circ$ (STEREO-A and -B, C2 scenario), respectively. For the last (C) event, we have also performed a fit with three viewpoints where all three spacecraft have a $120^\circ$ separation from each other (C3 scenario), and finally one fit with two viewpoints where only the height of the apex $h$ of the GCS model was altered. This latter fit will provide us with the opportunity to discuss the linear speed and associated errors derived from these two consecutive fits. In Figure~\ref{fig:syntheticGCS} the synthetic white-light images created for each of the GCS model events (listed in Table~\ref{tab:whitelightdetails}) are shown. Note that each event is Earth-directed, so that we have an (almost) halo CME in L1 data, but each CME is directed slightly differently. Hence, we should keep in mind when analysing the final data that, as we are also using different observation angles for multi-viewpoint fits, we cannot draw any conclusions about our fitting capacities between different viewpoints. We can, however, draw general conclusions about our fitting abilities, as will be presented in the next section.

\begin{table}[ht!]
    \centering
    \begin{tabular}{l|c|c|c}
         Configuration & A & B & C   \\
         \hline\hline
       Longitude $\phi$ [$^\circ$]& 25.7 & $-35.8$ & 20.1 \\
       Latitude  $\theta$ [$^\circ$] &11.7 & $-10.6$ & 2.5 \\        
    Axial tilt $\gamma$ [$^\circ$] &$-31.9$ &$-2.2$ & $-11.2$\\         
       Height of apex $h$ [R$_{\odot}$] & 13.91& 7.46& 8.69\\         
        Aspect ratio $\kappa$ & 0.29 & 0.32 & 0.47\\
        Half-angle $\alpha$ & 42.2 & 50.6 & 52.8\\
    \end{tabular}
    \caption{True GCS parameters for each of the configurations shown in Figure \ref{fig:config} and present in Section \ref{subsec:data-simple}}
    \label{tab:whitelightdetails}
\end{table}

\subsection{CME reconstructions of GCS synthetic white-light data} \label{subsec:synt_gcs}



\begin{figure*}[th!]
\centering
\includegraphics[width=.49\linewidth]{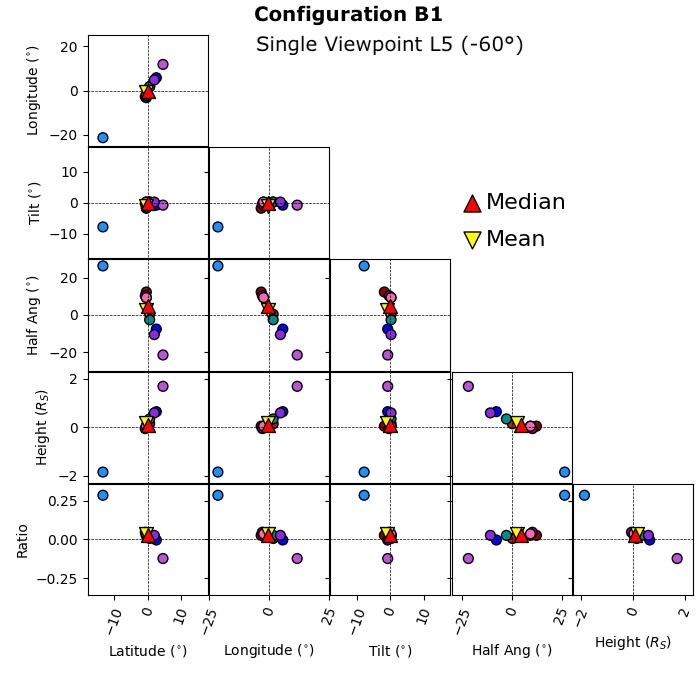}
\includegraphics[width=.49\linewidth]{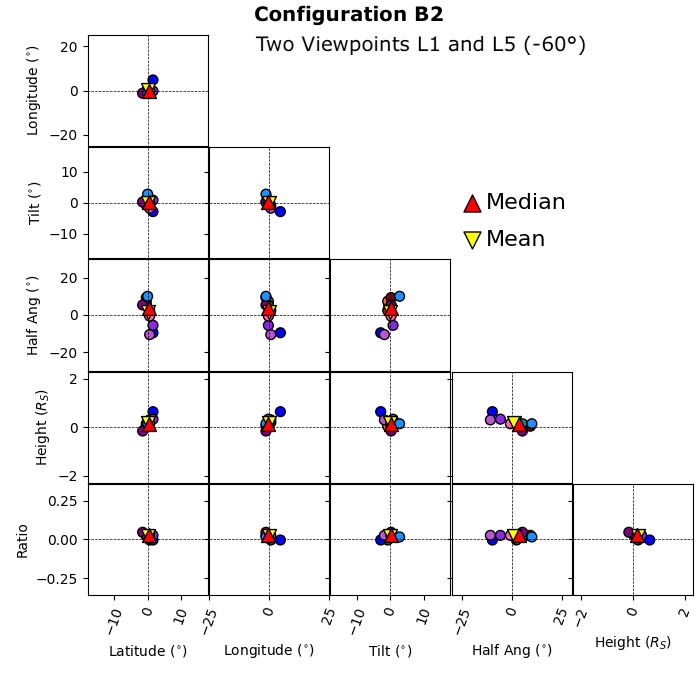}
\caption{Correlation plots of GCS CME parameters for the ten fits using synthetic white-light data from configuration B, two spacecraft separated by 60$^{\circ}$ (see Figure~\ref{fig:config}). Left - configuration B1: GCS CME parameters from fits made with only one viewpoint (L1).  Right - configuration B2: parameters from fits made with two viewpoints (L1 and L5). All of the values are shown relative to the true GCS fit on which the white-light image data is based. The dashed lines through the $(0,0)$ point show the true fit parameters. Each color represents a different user performing the fit, and the yellow downward and red upward triangles represent the mean and median of all fits, respectively. Please note the differing scales used for the various parameters.}
\label{fig:synt_L5}
\end{figure*}

Each of the seven configurations (three events A, B, and C and considering different number of spacecraft: A1, A2, B1, B2, C1, C2, and C3, as described in the previous section) have been fitted for a total of ten times (one for each member participating in this study). 
Since the true fit for the perfect GCS model fit is known, the results allow us to learn more about the minimum CME parameter uncertainties that should be taken into account when performing a GCS model fit on real observational data. In Figure~\ref{fig:synt_L5}, we present results for the event in configuration B in the form of a collection of correlation plots between all possible GCS fit parameters. On the left side (B1), we show results for fittings of one viewpoint (L1) and on the right side (B2) for fittings of two viewpoints (L1 and L5, with $60^\circ$ separation). All of the values are shown relative to the true GCS fit on which the white-light image data is based. The dashed lines through the $(0,0)$ point show the true fit parameters. Each color represents a different user performing the fit, and the yellow downward and red upward triangles represent the mean and median of all fits, respectively. The general conclusions drawn from below based on the event in configuration B are also valid for all of the other configurations, hence, we show only results for this particular event. All fittings and results are provided as supplemental material. 

\begin{figure*}[ht]
\centering
\includegraphics[width=.3\linewidth]{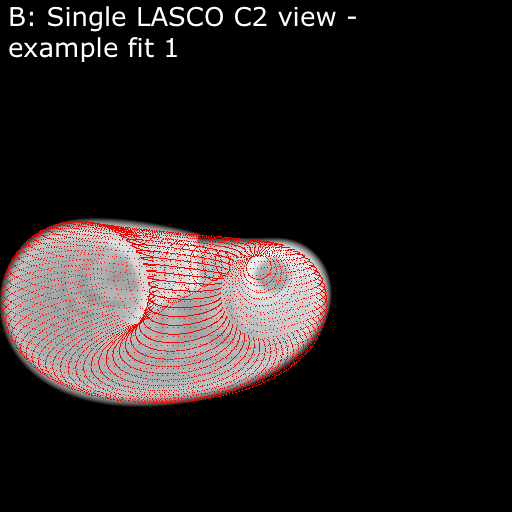}
\includegraphics[width=.3\linewidth]{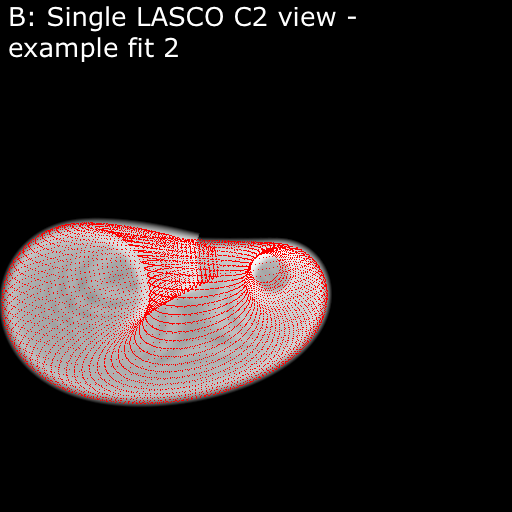}
\includegraphics[width=.3\linewidth]{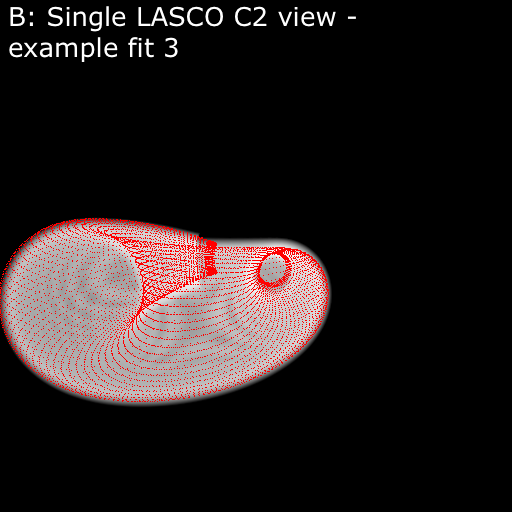}
\caption{Examples of different fits to LASCO synthetic images using only one viewpoint from configuration B (L5-like), performed by three different observers. These examples illustrate a wide range in values for the half-angle. All three fits visually appear equally valid, however, they correspond to half-angles of $29^\circ$, $48^\circ$, and $63^\circ$, and to latitudes of $-6^\circ$, $-10^\circ$, and $-11^\circ$ from left to right, demonstrating the projection effects.}
\label{fig:width_example}
\end{figure*}

As can be seen from Figure~\ref{fig:synt_L5}, when we move from one viewpoint (left side) to two viewpoints (right side), the performed fits tighten around the true values. For the one-viewpoint fittings, we can observe many correlations, e.g., those who fit a higher latitude, also tend to fit a smaller half-angle $\alpha$. Such correlations appear as diagonal lines in the plots. We note that this is a result of the specific technique chosen (GCS) in combination with projection effects, as we are only considering one viewpoint. As a result, the correlation can change or invert depending on the propagation direction and tilt of the CME and/or the location of the spacecraft compared to the CME. In Figure~\ref{fig:width_example}, we present three fittings that have been performed on configuration B with one viewpoint. All three fits visually appear as a good fit, however, they correspond to half-angles of $29^\circ$, $48^\circ$, and $63^\circ$, and to latitudes of $-6^\circ$, $-10^\circ$ and $-11^\circ$ from left to right, demonstrating the projection effects. The main difference between the fits and the true parameters are related to fitting the appearance of the CME ``legs" in the synthetic GCS white-light image. However, an observer performing GCS fits on actual white-light CME data would not see structured CME ``legs" to visibly distinguish the fits and, as such, all three fits would be equally valid. For two spacecraft (see Figure~\ref{fig:synt_L5} right), the correlations almost entirely disappear. Only a small correlation is left for the half-angle and height. While having two viewpoints still results in fitting errors, the errors are significantly smaller and systematic trends in those errors are removed.

\begin{figure*}[ht!]
\centering
\includegraphics[width=.32\linewidth]{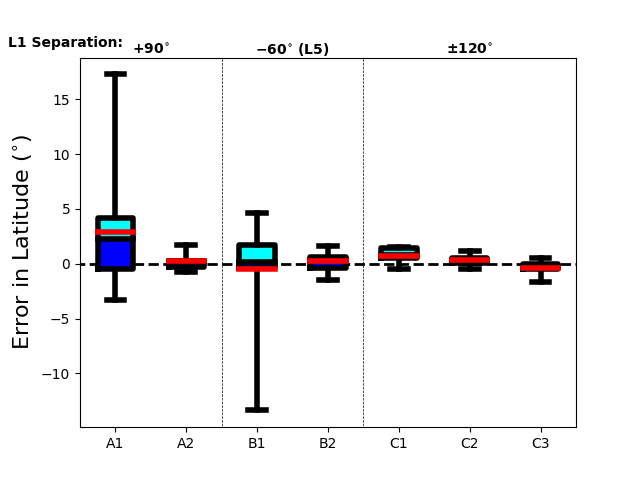}
\includegraphics[width=.32\linewidth]{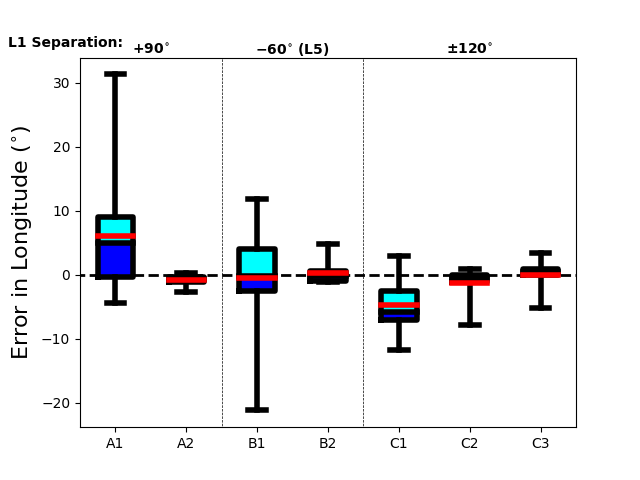}
\includegraphics[width=.32\linewidth]{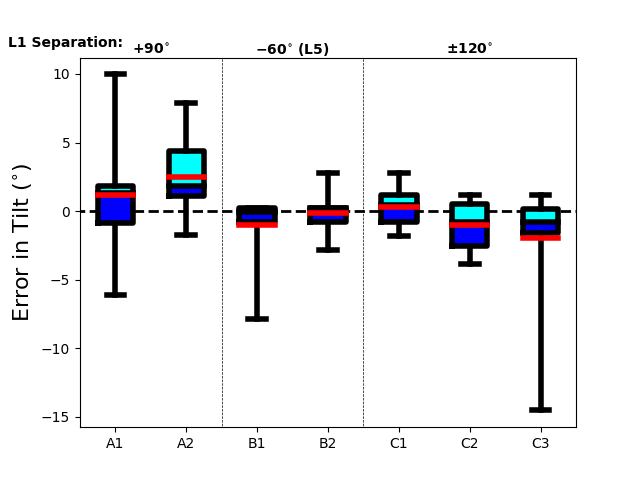}\\
\includegraphics[width=.32\linewidth]{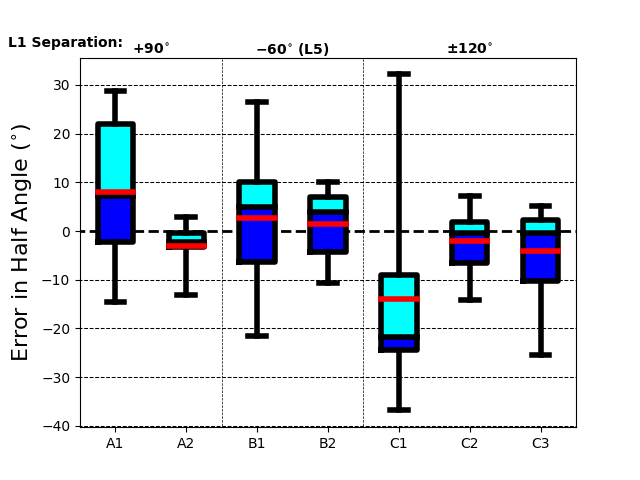}
\includegraphics[width=.32\linewidth]{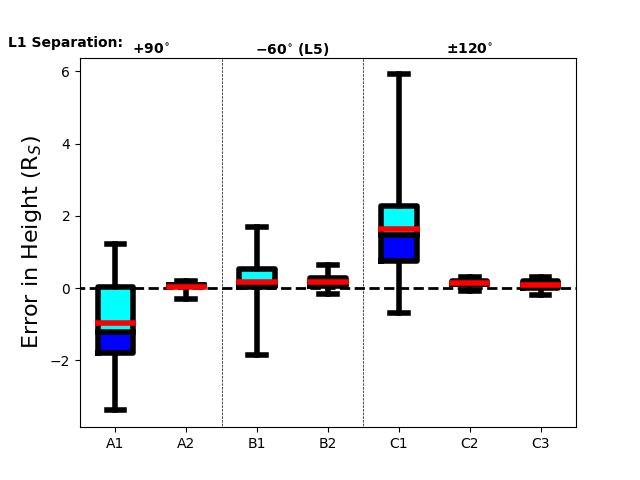}
\includegraphics[width=.32\linewidth]{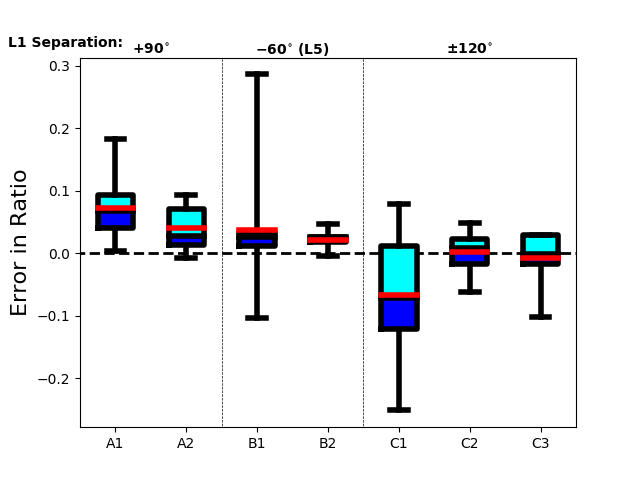}\\
\includegraphics[width=.32\linewidth]{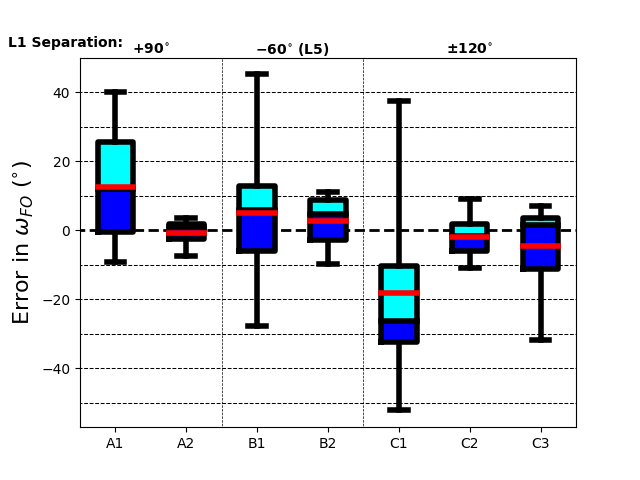}
\includegraphics[width=.32\linewidth]{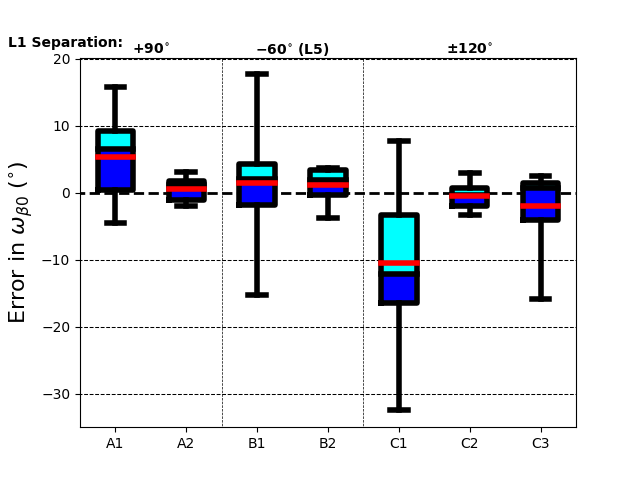}
\includegraphics[width=.32\linewidth]{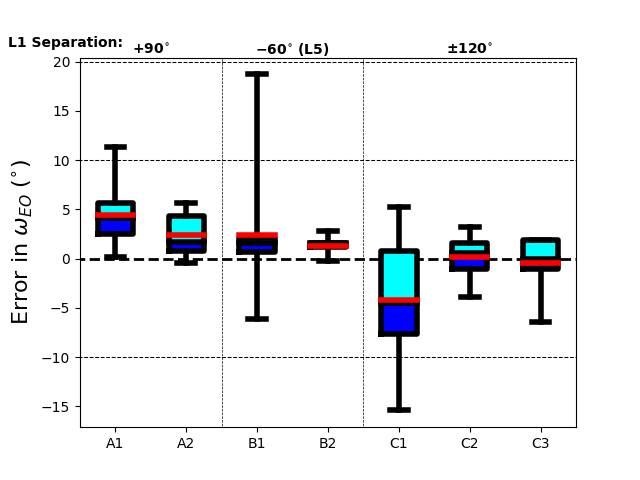}\\
\includegraphics[width=.32\linewidth]{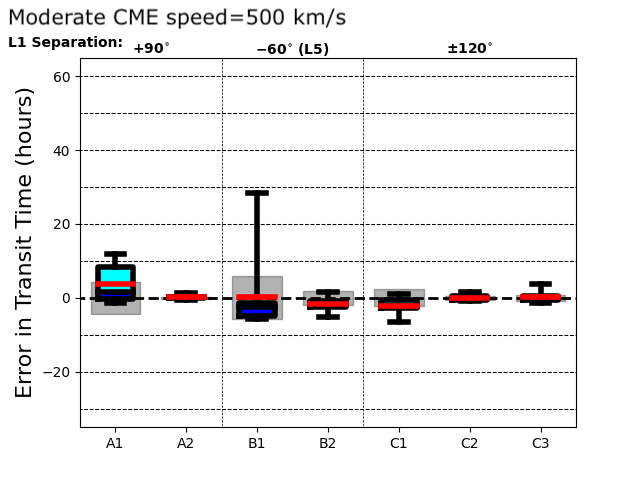}
\includegraphics[width=.32\linewidth]{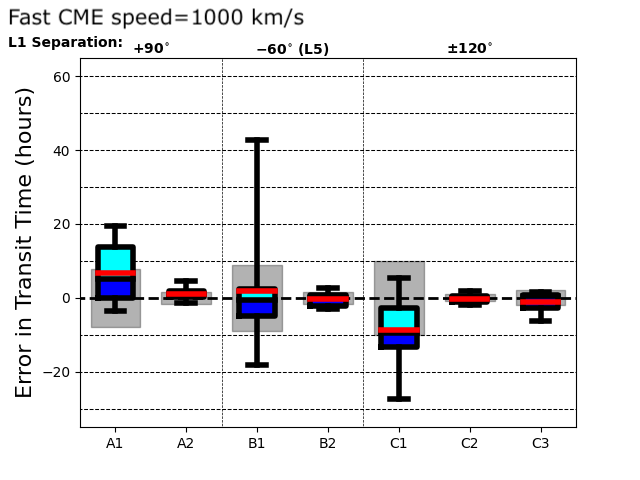}
\includegraphics[width=.32\linewidth]{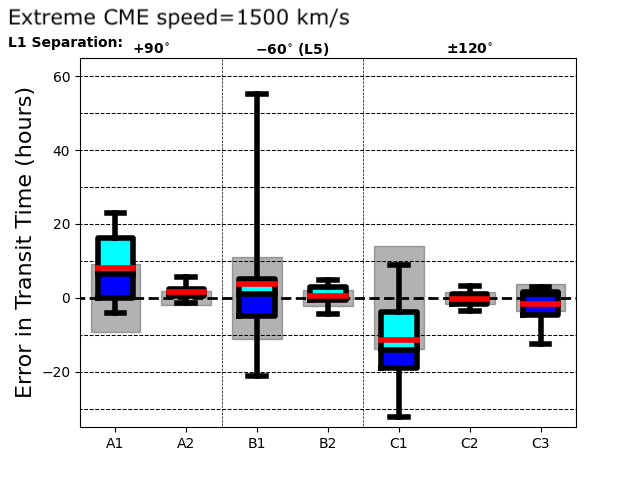}
\caption{Box and whisker plots of GCS fit parameters for ten fits relative to the true GCS fit for three events in configurations A, B, and C (see Figure~\ref{fig:config}).  The median and mean are shown by the black and red horizontal lines respectively, the box represents the first and third quartiles, and the whiskers show the full range. The gray boxes on the bottom row are representative of the MAE of the transit time, which are listed in Table \ref{tab:anteatr}. Rows 1-2: GCS parameters of latitude, longitude, axial tilt, half angle, height of apex, aspect ratio $\kappa$ (direct GCS outputs). Row 3: face-on width, face-on width at $\beta=0^\circ$, edge-on width (derived parameters). For rows 1--3, please note the differing scales used for the y-axes. Row 4: arrival time errors from the ANTEATR model for moderate (left), fast (middle), and extreme (right) hypothetical CME speeds. The configurations are labeled on the x-axis such that 1: single L1 spacecraft, 2: L1 and one additional spacecraft, 3: L1 and two additional spacecraft. The spacecraft separation angle with respect to L1 is annotated above each plot (+90$^{\circ}$, -60$^{\circ}$, and $\pm$120$^{\circ}$ for configurations A, B and C respectively).}
\label{fig:box}
\end{figure*}

The fitting errors are further explored in Figure~\ref{fig:box} in the form of box and whisker plots of the GCS fit parameters for ten fits relative to the ``true" GCS parameters for three CME events A, B, and C (see Figure~\ref{fig:config}).  The median and mean are shown by the black and red horizontal lines, respectively. The box represents the first and third quartiles, and the whiskers show the full range. The first and second rows show the GCS parameters of latitude, longitude, axial tilt, half angle, height of apex and aspect ratio $\kappa$ (direct GCS outputs). In the third row, we show the derived parameters of the face-on width $\omega_{\mathrm{FO}}$, the face-on width at $\beta=0^{\circ}$ ($\omega_{\mathrm{B0}}$, and the edge-on width $\omega_{\mathrm{EO}}$ (see Section \ref{subsec:gcs}). We discuss the fourth row of figures in Section \ref{subsec:arrivaltimesynthetic}. The configurations are labeled on the x-axis by the number of spacecraft used to fit the CME, 1: single L1 spacecraft, 2: L1 and one additional spacecraft, 3: L1 and two additional spacecraft. The spacecraft separation angle with respect to L1 is annotated above each plot (+90$^{\circ}$, $-$60$^{\circ}$, and $\pm$120$^{\circ}$ for events A, B and C respectively).  Similar to the correlation plots in Figure \ref{fig:synt_L5}, we conclude that errors decrease and the box and whiskers (range) move closer to the true value (error of 0) when going from fitting only a single L1 spacecraft viewpoint (A1, B1, C1) to adding one (A2, B2, C2) or two additional (C3) spacecraft.  Particularly for event C, going from two to three spacecraft does not result in a significant change in fitting error. This result is valid for this particular event and may not apply to other configurations, but nevertheless this hints towards the crucial need for two viewpoints for reducing CME parameter errors used for CME arrival forecasting. One exception to the clear trend in error decrease is with the axial tilt. For events A and C, the tilt error slightly increases with the number of spacecraft, while it remains unchanged for event B. Using the definitions for the face-on and edge-on widths by \citet{thernisien2011}, $\omega_{\mathrm{FO}}=2(\alpha+\delta)$ and $\omega_{\mathrm{EO}}=2\delta$, respectively, one can easily show using error propagation that the corresponding width errors are $\Delta\omega_{\mathrm{FO}}=2\sqrt{\Delta\alpha
^{2}+\Delta\delta^{2}}$ and $\Delta\omega_{\mathrm{EO}}=2\Delta\delta$, respectively. The spread of the edge-on width will thus only depend on the spread of the aspect ratio $\kappa = \sin \delta$, while the spread of the face-on width (and similarly, the face-on width at $\beta=0^\circ$) will also depend on the half angle $\alpha$. Since the error spread in the half angle is much larger compared to the error spread in the aspect ratio, this will result in a much larger error spread of the face-on width compared to the edge-on width as is visible in Figure \ref{fig:box}.

\begin{figure}[p]
\centering
\includegraphics[width=1.0\linewidth]{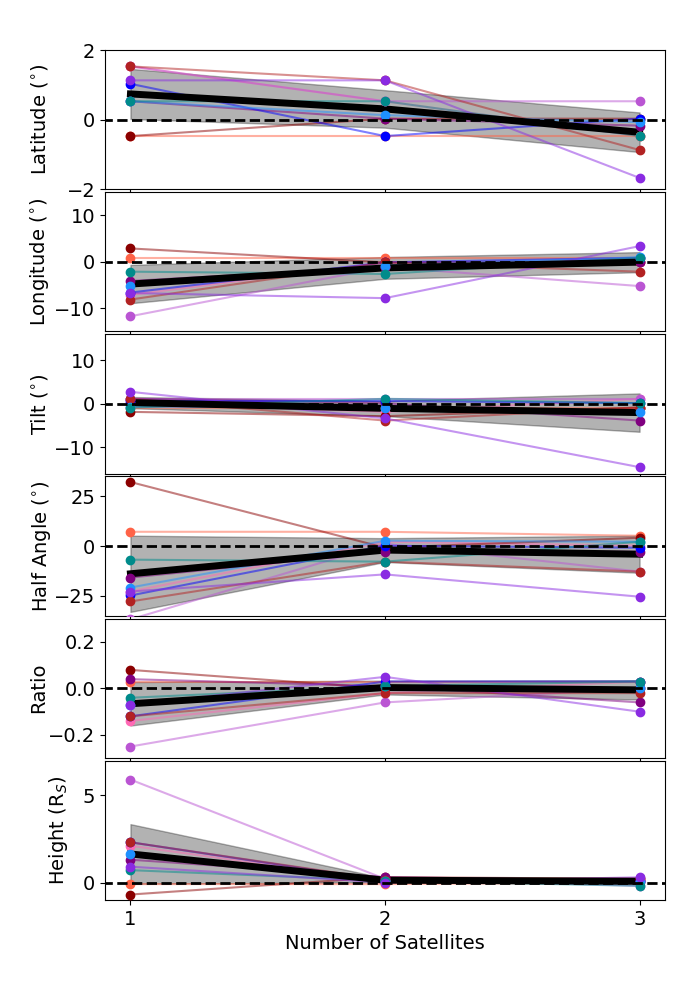}
\caption{The relative error in reproducing the true GCS fit parameters (latitude, longitude, tilt, half angle, and ratio) for ten different fits for event C, with one (L1), two (+120$^{\circ}$), or three spacecraft (L1 and $\pm$120$^{\circ}$). Each color represents an individual fit, the solid black line shows the mean, and the grey shaded region shows one standard deviation about the mean. The mean and standard deviations values plotted in this figure are shown in Table \ref{tab:mae_synth}.}
\label{fig:synthline}
\end{figure}

Figure~\ref{fig:synthline} shows the relative error in reproducing the true GCS fit parameters (latitude, longitude, axial tilt, half angle, and aspect ratio) for the ten different fits for event C with one (L1), two (L1 and +120$^{\circ}$), or three spacecraft (L1 and $\pm$120$^{\circ}$). Each color represents an individual fit, the solid black line shows the mean, and the grey shaded region shows one standard deviation about the mean. In general, as the number of spacecraft used in the fit increases, the fits become more accurate and the range in errors decrease. Note that the error in the latitude does not decrease, but neither does it increase and the latitudinal spread is the smallest of all parameters.  In general, latitude is the most straightforward parameter to fit because it is the least impacted by projection effects, as long as the CME is not too high in latitude. However slower CMEs can show a larger deviation from the source region (including in latitude), which results in a non-radial propagation, that the GCS method cannot cope with \citep[e.g.,][]{temmer2009}. Again, the tilt is the exception which begins with a surprisingly good single spacecraft fit. For this particular event, the tilt may have been straightforward to fit with only a single spacecraft (C1) viewpoint because the true tilt was nearly $0^\circ$ and the CME was face-on toward the observer. As more spacecraft views are added, the observer is faced with more options for the tilt that can also fit the other viewpoints.

In addition to the graphical exploration of fitting errors in the earlier figures, we present the standard deviation (SD) and Mean Absolute Error (MAE) averaged over different sets of synthetic GCS configurations in Table \ref{tab:mae_synth}. Column 1 is averaged over all synthetic GCS configurations (A1 through C3), and we also show the minimum and maximum MAE. Column 2 is averaging over single spacecraft configurations A1, B1, and C1. Column 3 is averaging over two spacecraft configurations A2, B2, and C3 and column is the three spacecraft configuration C3.  The trends in this table are also clearly visible in Figure \ref{fig:synthline} as the solid black line (mean) and grey shaded regions (one standard deviation).  We note that overall the spread in latitude is lower than the one in longitude, consistent with the analysis from \citet{thernisien2009}. Furthermore, the spread in half-angle is significantly larger than in the other parameters, especially important considering that the half-angle is a crucial parameter for the forecasting of CME arrival times (see Section~\ref{subsec:arrivaltimesynthetic}). The maximum MAE values are all from the single-viewpoint configurations A1 and C1, while the minimum MAE values are from a mix of the two- and three-viewpoint configurations, i.e. A2, B2, and C3. The averaged standard deviations (spread of the fits) and MAE are largest for the CME half-angle and longitude.

\begin{table*}[th!]
    \centering
    \begin{tabular}{l|r|r|r|r|r|r|r|r|r|r|}
         & \multicolumn{4}{c|}{All Viewpoints} & \multicolumn{2}{c|}{One Viewpoint} & \multicolumn{2}{c|}{Two Viewpoints}  & \multicolumn{2}{c|}{Three Viewpoints}  \\
         Synthetic GCS 
        & \multicolumn{4}{c|}{A1 -- C3} & \multicolumn{2}{c|}{A1, B1, C1} & \multicolumn{2}{c|}{A2, B2, C2}  & \multicolumn{2}{c|}{C3}  \\ \hline
         & Average & \multicolumn{3}{c|}{MAE} & \multicolumn{2}{c|}{Average} & \multicolumn{2}{c|}{Average} &  & \\
         & \multicolumn{1}{c|}{$\sigma$} &  \multicolumn{1}{c|}{Avg} & \multicolumn{1}{c|}{Min}  & \multicolumn{1}{c|}{Max} & \multicolumn{1}{c|}{$\sigma$}  & \multicolumn{1}{c|}{MAE} & \multicolumn{1}{c|}{$\sigma$} & \multicolumn{1}{c|}{MAE}  & \multicolumn{1}{c|}{$\sigma$} & \multicolumn{1}{c|}{MAE}\\
         \hline\hline
Latitude [$^\circ$] 	&	1.95	&	1.45	&	0.48	&	4.28	&	3.63	&	2.59	&	0.73	&	0.62	&	0.57	&	0.48	\\
Longitude [$^\circ$] 	&	4.17	&	3.44	&	0.98	&	7.77	&	7.37	&	6.36	&	1.65	&	1.19	&	2.16	&	1.47	\\
Tilt [$^\circ$] 	&	2.62	&	2.01	&	1.07	&	3.3	&	2.64	&	1.9	&	2.02	&	1.94	&	4.41	&	2.53	\\
Half angle ($\alpha$) [$^\circ$] 	&	10.6	&	9.87	&	4.21	&	21.8	&	15.52	&	15.56	&	6.11	&	5.19	&	9.33	&	6.86	\\
Height [$R_{\odot}$] 	&	0.64	&	0.61	&	0.12	&	1.8	&	1.28	&	1.21	&	0.15	&	0.16	&	0.16	&	0.15	\\
Ratio ($\kappa$) 	&	0.051	&	0.049	&	0.023	&	0.096	&	0.078	&	0.076	&	0.027	&	0.03	&	0.042	&	0.03	\\
$\omega_{\mathrm{EO}}/2 = \delta = \arcsin\kappa$ 	&	2.92	&	2.81	&	1.32	&	5.51	&	4.47	&	4.36	&	1.55	&	1.72	&	2.41	&	1.72	\\
$\omega_{\mathrm{FO}}/2 =\alpha + \delta$ 	&	13.52	&	12.68	&	5.53	&	27.31	&	19.99	&	19.92	&	7.66	&	6.91	&	11.74	&	8.58	\\
$\omega_{\mathrm{FO}(\beta=0^\circ)}/2$ 	&	6.65	&	6.25	&	2.76	&	12.77	&	9.65	&	9.61	&	3.81	&	3.44	&	5.79	&	4.26	\\
    \end{tabular}
    \caption{Mean absolute errors (MAE) and standard deviations ($\sigma$) for each GCS fit parameter from the synthetic GCS white-light images.  Column 1: averaged over all synthetic GCS configurations (A1 through C3), and we also show the minimum and maximum MAE. Column 2: standard deviations and MAE averaged over single spacecraft configurations A1, B1, and C1. Column 3: standard deviations and MAE averaged over two spacecraft configurations A2, B2, and C3. Column 4: standard deviations and MAE for the three spacecraft configuration C3.  The trends in this table are also clearly visible in Figure \ref{fig:synthline} as the solid black line (mean) and grey shaded regions (one standard deviation).}
\label{tab:mae_synth}
\end{table*}

\subsection{Derived speed from GCS fits}

When fitting real observational data, two (or more) consecutive GCS fits at different times are usually performed. Using the change in the height of the apex and the time difference between the observations allows one to determine the CME speed profile. In this study, GCS fits were only performed for a single height for all configurations except for configuration C2 (L1 and 120$^\circ$ STEREO-A) for which we performed an extra fit where only the height of the CME was altered (see Figure \ref{fig:syntheticGCS}). However, since the fits were performed on synthetic images, there are no real times associated with them. Still, we wish to explore how errors in height measurements propagate to speed errors for a range of CME speeds. Therefore, we created a range of hypothetical times differences that correspond to a range of CME speeds from 400 to 2000 km~s$^{-1}$ with increments of 400 km~s$^{-1}$, based on the `true' heights of 8.69 and 12.68 R$_{\odot}$. We can then calculate the linear CME speeds obtained from the fits and compare to the true CME speeds derived from using the true heights. We determine the mean error (ME), the MAE, as well as the root mean squared error (RMSE). For all of the considered times differences and corresponding speeds, the ME are always negative, ranging between -3 and -16 km~s$^{-1}$ from low to high considered CME speeds, meaning that we are always underestimating the CME speed. One should keep in mind that this may be due to the specific way the synthetic white light image data is generated as well as to the specific parts in this image that are measured by each observer, but we expect that these variations will increase even more when considering real observational data. MAE range from 10 to 49 km~s$^{-1}$ (3\% error), with corresponding root mean square errors from 14 to 69 km~s$^{-1}$.  Note that this experiment represents a lower bound estimate of the errors that arise from determining the CME speed from coronagraph images.

\subsection{CME arrival time using the parameters of the GCS fittings}\label{subsec:arrivaltimesynthetic}

Lastly, we consider the propagation of the GCS fitting errors from the synthetic images to CME arrival time errors of the leading edge at 1 au using the ANTEATR model \citep[see][for a model description]{kay2018,kay2020}. GCS fit parameters of latitude, longitude and width $\omega_{\mathrm{FO}(\beta=0^\circ)}$ were used as input to the ANTEATR ensemble model (therefore ten members in the ensemble relating to the ten performed fits). A separate ensemble was constructed for each configuration (A1 to C3: 7 total) and for three different hypothetical CME speeds, for a total of 21 ensembles. Since GCS fits were only performed for a single time for nearly all configurations, GCS derived speeds are not available. Therefore we consider three CME inputs speeds that are moderate (500 km~s$^{-1}$), fast (1000 km~s$^{-1}$), and extreme (1500 km~s$^{-1}$). Note that these do not correspond to the speeds and related errors discussed in the previous section. In Table \ref{tab:anteatr} we list the CME arrival time MAE from the ANTEATR model ensemble results for each configuration and speed (total of 21 options). Row 4 of Figure \ref{fig:box} shows the arrival time errors from the ensemble for moderate 500 km~s$^{-1}$ (left), fast 1000 km~s$^{-1}$ (middle), and extreme 1500 km~s$^{-1}$ (right) hypothetical CME speeds, where we have kept the y-axis scale identical for all three plots. The gray boxes are representative of the MAE for each configuration, which are listed in Table \ref{tab:anteatr}. We can examine these errors by the number of viewpoints used: single (A1, B1, C1), two (A2, B2, C2), and three (C3). For the single viewpoint fits (A1, B1, C1), the modeling results show a mean absolute arrival time error of 4, 9, and 11 hours for moderate, fast, and extreme speed CMEs respectively. From Figure \ref{fig:synt_L5} we observed that the CME position is fairly accurate compared to the half angle and ratio, therefore the range in arrival time errors arise primarily from the range of CME widths which lead the CMEs to experience drag differently within the model. For moderate speed CMEs the arrival time errors of the ten fits (relative to the true fit) have a smaller range than faster CMEs. This is because the drag force in the model is less effective for moderate speed CMEs. The drag force strength experienced by the faster CMEs is sensitive to the CME width. The ensemble CME parameter sensitivity study of \cite{kay2020} found that an accuracy of $5^\circ-10^\circ$ in CME width (in addition to other parameters) is necessary to achieve a CME arrival time error of 5 hours or less (for fast and extreme CMEs). We can see that the trend of error decreasing with increasing spacecraft viewpoints also holds true for the arrival time error. For the two viewpoint fits (A2, B2, C2), the average of the MAE for moderate, fast, and extreme speed CMEs is 1.0, 1.4, and 1.9 hours respectively. This is comparable to the three viewpoint fits (C3), which have a MAE of 0.9, 2.1, and 3.7 hours. Again, we note that this experiment represents a lower bound estimate of the arrival time errors that arise from determining the CME speed from coronagraph images. This calls for the critical need to have at least two viewpoint observations of CMEs to reduce CME arrival time errors arising from CME measurements. 

\begin{table}[htb]
    \centering
    \begin{tabular}{r|r|r|r||r|r|r||r}
         \multicolumn{1}{c|}{Speed} & \multicolumn{7}{c}{CME Arrival Time MAE (hours)}\\
         (km~s$^{-1}$) & A1 & B1 & C1 & A2 & B2 & C2 & C3  \\
         \hline\hline
500 & 4.3 & 5.9 & 2.3 & 0.4 & 2.0 & 0.6 & 0.9 \\
1000 & 7.8 & 8.8 & 10.0 & 1.6 & 1.6 & 1.0 & 2.1 \\
1500 & 9.2 & 11.2 & 14.0 & 2.0 & 2.2 & 1.6 & 3.7 \\
         \hline
    \end{tabular}
    \caption{CME arrival time Mean Absolute Error (MAE) from the ANTEATR model for each configuration and speed.}
    \label{tab:anteatr}
\end{table}


\section{Synthetic white-light data derived from an MHD Simulation} \label{sec:mhd}

\subsection{White-light images derived from an MHD simulation}

For the second part of our study, to increase the complexity of our synthetic images we generated simualted white-light data from an idealized thermodynamic MHD simulation of a hypothetical fast CME, which was initiated from a stable pre-eruptive magnetic state.  This allows us to create a situation that is closer to observational data but where we are still able to derive a ``true'' set of CME parameters from the full 3D simulation data available, which is not possible to obtain for actual CME observations. The simulation and its application to the study of Solar Energetic Particles (SEPs) has been described elsewhere \citep{schwadron2015a,schwadron2015b,schwadron2017}. 

The simulation was performed using the MAS code \citep{riley2012,riley2019,torok2018}, which advances the standard viscous and resistive MHD equations forward in time in a spherical coordinate system. Radiative losses, thermal conduction parallel to the magnetic field, and an empirical coronal heating function are included. We simulate the region from the solar surface to 20\,$R_{\odot}$ with $251 \times 301 \times 261$ grid cells in the $r$, $\theta$, and $\phi$ directions. The grid is non-uniform, with more points to capture variations at lower altitudes and around the active region where the CME originates. 

\begin{figure*}[ht]
\centering
\includegraphics[width=.9\linewidth]{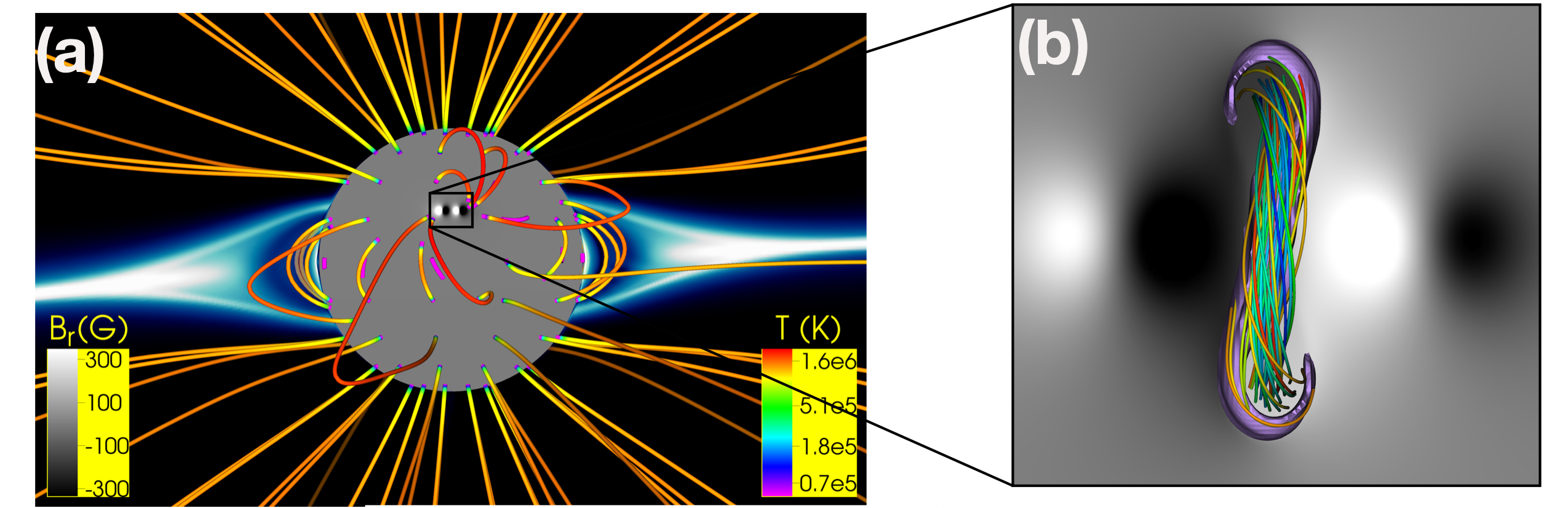}
\caption{Summary of MHD CME simulation setup. (a) A selection of field lines for the background field (coloured according to their temperature), overlaid on a simulated white-light image, where the solar surface is coloured according to the polarity of the magnetic field.  (b) Flux configuration that was embedded into the background field shown in (a).}
\label{fig:mhdsetup}
\end{figure*}

We chose this particular simulation for its simplicity: an initial magnetic configuration consisting of one active region and a global background dipole field, resembling solar minimum conditions. After relaxing the configuration to an equilibrium state, which included a reasonably realistic two-speed solar wind, we inserted a flux rope in magnetic equilibrium along the polarity inversion line of the active region, using the technique developed by \citet{titov2014}. After a further relaxation phase, during which the system adjusted to this modification, we triggered the eruption of the flux rope by imposing slow, localized converging flows at the lower boundary. In Figure~\ref{fig:mhdsetup}, we show the active region and inserted flux rope. Figure~\ref{fig:mhdsetup}(a) shows a selection of field lines for the background field (coloured according to their temperature), overlaid on a simulated white-light image, where the solar surface is shaded according to the polarity of the magnetic field. Figure~\ref{fig:mhdsetup}(b) shows the flux configuration that is embedded into the background field shown in (a). The resulting CME rapidly accelerates to more than 3000~km~s$^{-1}$ at low coronal heights, after which it quickly slows down and propagates further with a nearly-constant speed of about 1000~km~s$^{-1}$ before it reaches 3\,$R_{\odot}$. The interplanetary evolution of this event is explored in more detail by \citet{lionello2013}. Fits are performed about 36 minutes after the simulation start time, when the CME is approximately at a heliocentric distance of 3.6\,$R{_\odot}$, after the CME has reached the constant speed phase. Note that the simulation is of a hypothetical CME and does not correspond to an actual event.

\begin{figure*}[ht]
\centering
\includegraphics[width=.95\linewidth]{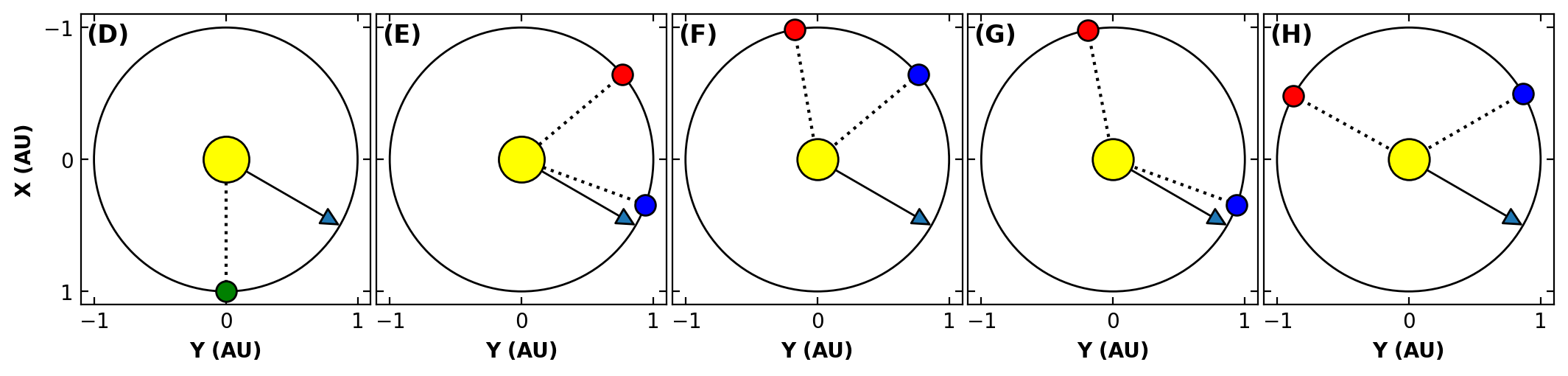}
\caption{Spacecraft configurations and CME directions used for creating MHD simulated white-light images of the CME in Heliocentric Earth Ecliptic (HEE) coordinates. The arrow denotes the propagation direction of the CME in the ecliptic plane. Events E and F have a $60^\circ$ separation, while events G and H have a $120^\circ$ separation.}
\label{fig:simconfigmhd}
\end{figure*}

\begin{table*}[ht]
    \centering
    \begin{tabular}{l|c|c|c|c}
         Event & \centering{Viewpoint 1 [$^\circ$]} & STEREO-B Header Date & \centering{Viewpoint 2 [$^\circ$]}   & STEREO-A Header Date \\
         \hline\hline
      D & \;90 & 2008-07-09T10:38:28.360 & - & - \\
      E & 160 & 2008-07-09T10:38:28.360 & 220 & 2008-07-09T10:37:30.004 \\
      F & 220 & 2008-07-09T10:38:28.360 & 280 & 2008-07-09T10:37:30.004 \\
      G & 160 & 2009-10-16T10:39:49.043 & 281 & 2009-10-16T10:39:00.005 \\
      H & 210 & 2009-10-16T10:39:49.043 & 331 & 2009-10-16T10:39:00.005 \\
    \end{tabular}
    \caption{Carrington longitude locations of the spacecraft viewpoints for the MHD simulated white-light data as shown in Figure \ref{fig:simconfigmhd}. The``DATE-OBS" from observational data headers from STEREO-A and STEREO-B which were applied to the MHD simulated white-light images are also listed.}
    \label{tab:sim}
\end{table*}

We have generated simulated white-light images for five different spacecraft configurations (labelled D through H), using only the data of the CME simulation described above. The simulated white-light images were created using a ray-tracing technique through the MHD density volume, taking into account the effects of Thomson scattering present in actual observational white-light data. For each configuration, the same CME is thus observed from different viewpoints. However, we note that the users performing the CME reconstructions did not know they were fitting the same CME observed from different perspectives, but were told to fit five separate ``events". Also, those involved in preparing the simulated white-light data did not perform any reconstructions. In Table \ref{tab:sim} and Figure \ref{fig:simconfigmhd} the exact details of the spacecraft configurations can be found. 

Configuration D has one viewpoint and the other four (E--H) all have two viewpoints. Of the configurations/events with two viewpoints, events E and F have a $60^\circ$ separation, and events G and H have a $120^\circ$ separation. While not listed in Table~\ref{tab:sim}, we want to note that the latitude and the radius of the observing spacecraft locations also slightly change. This is because actual observational data headers from STEREO-A and STEREO-B were applied to the MHD simulated white-light images, which allowed us to appropriately use the SolarSoft GCS program. Note that due to a STEREO data gap during the $120^\circ$ separation, the STEREO headers corresponding to a $121^\circ$ separation were used. Therefore, the locations correspond to the real locations of the spacecraft at a given time corresponding to the desired spacecraft longitudinal separation. As such, latitude and radial distance also correspond to the spacecraft locations as well as the changing field of view as a result of the changing radial distance. These details are taken into account when producing the simulated white-light images. 

Unlike for the synthetic white-light images generated from the GCS model, we do not have an obvious ground truth directly available from the simulation. As the CME travels through the domain of the simulation, it changes its kinematics. Moreover, the overall morphology of the simulated CME does not straightforwardly compare to that of the GCS hollow croissant, implying that the application of the GCS technique will not be a perfect match to the simulated CME shape. However, because we have the full 3D simulation data available compared to just a few 2D viewpoints for actual observational data, we are able to derive a ground-truth value. We briefly discuss how we arrived at the final truth value and uncertainty that was used to compare against the fits to draw conclusions.

\begin{figure}[ht]
\centering
\includegraphics[width=1.0\linewidth]{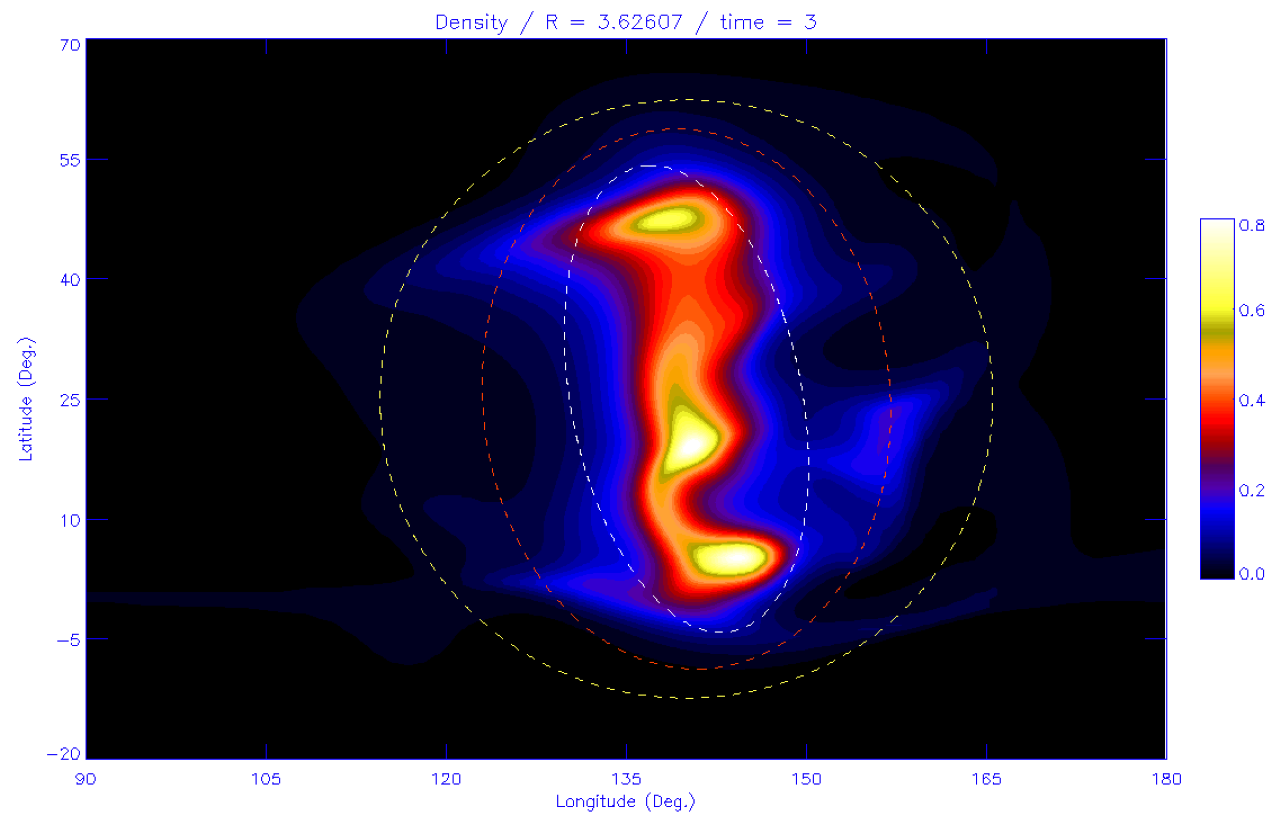}
\caption{Figure showing the MHD scaled density at R=3.62 $R_{\odot}$ spherical slice in longitude and latitude at the same time step of the simulated white-light data (t =3 code time units).  R=3.62 $R_{\odot}$ is the slice at which the CME cross-section had the largest extent in both latitude and longitude.  The medium red ellipse overlay was used to derive the ground truth for the CME parameters of longitude, latitude, tilt, half-angle and ratio, and the white ellipses were used as the uncertainty.}
\label{fig:simfigure}
\end{figure}

To determine the ground-truth GCS parameters that fit the simulated CME best, we thoroughly explored how to track and detect the CME within the full simulation volume using different automated and manual methods. The automated method involved tracking the maximum scaled CME density through out the volume, and detecting the CME edges in 2D slices. However, after several tests we have decided to deduce the ground truth in the following way, using a manual method. We used the simulation data time step corresponding to the simulated white-light data that was provided to the observers (corresponding to simulation time $t = 3$ --code time units-- about 36 minutes after simulation start).  First, to derive the true height of the CME we reviewed a series of radial spherical slices (longitude versus latitude) of the MHD density, scaled by the radial distance $r^2$ (an example slice is shown in Figure \ref{fig:simfigure}). We searched for the radial distance (height) at which the CME front/nose appeared in the density slices. The full team unanimously agreed that the CME was visible in four consecutive heights (3.99, 4.06, 4.13, 4.20 $R_{\odot}$). The CME front was not visible after 4.20 $R_{\odot}$ while it was definitely visible at the lower height of 3.99 $R_{\odot}$. Therefore we chose 4.13\,$R_{\odot}$ as the ground truth CME height, and we used the lower and higher heights to create uncertainties on this ground truth value: $4.13^{+0.07}_{-0.14}$ $R_{\odot}$. Next, we derived the true values of the CME longitude, latitude, tilt, and size by using the spherical slice at which the CME cross-section had the largest extent in both latitude and longitude (at time step $t = 3$), which was $R = 3.62\,R_{\odot}$. Figure \ref{fig:simfigure} shows the MHD scaled density at the $R = 3.62\,R_{\odot}$ spherical slice in latitude and Carrington longitude. At this height, we have manually fit three ellipses to the CME cross-section in the density plot that the team agreed reasonably enveloped the enhanced density region corresponding to the CME. The smallest (white) ellipse envelopes only the highest density of the CME, excluding the lower density parts of the CME, while the largest (yellow) ellipse, envelopes the maximum full extent of the CME density. Finally, the team agreed that the medium (red) ellipse which envelopes most of the CME density is the most representative ellipse fit to be used as the truth. The small and large (white and yellow) ellipses were then used to derive uncertainties. From these ellipses, we derived the MHD ground truth and uncertainties for the latitude, longitude, and tilt angle, using the center and orientation of the ellipse. In addition, the GCS half-angle $\alpha$ and ratio $\kappa$ were derived from the ellipse minor and major axes. Note that the ellipse size was used to compute the non-angular GCS widths in \citep[${W_\mathrm{FO}}$ and ${W_\mathrm{EO}}$ in][]{thernisien2011}, from which the half-angle and ratio were then derived. We present the MHD ground truth GCS CME parameters and their uncertainties in Table~\ref{tab:mhdtruth}. Since the small, medium, and large ellipse fits were performed independently, the resulting uncertainties are not symmetric.

\begin{table*}[th!]
    \centering
    \begin{tabular}{cccccc}
    \hline
        Latitude [$^\circ$] & Longitude [$^\circ$] & Tilt [$^\circ$] & Half-angle ($\alpha$) [$^\circ$] & Height [$R_{\odot}$] & Ratio ($\kappa$)\\
    \hline\hline\\
    \vspace{2mm}
    26.07$^{+2.67}_{-0.56}$ & 140.06$^{+3.19}_{-0.00}$ & -88.09$^{+4.43}_{-1.91}$ & 23.1$^{+1.20}_{-4.00}$ & 4.13$^{+0.07}_{-0.14}$ & 0.305$^{+0.172}_{-0.189}$\\
    \hline
    \end{tabular}    
\caption{Derived ground ``truth" GCS CME parameters from the MHD simulation of the CME, and their uncertainties.}
\label{tab:mhdtruth}
\end{table*}

\begin{figure}[p]
\centering
\includegraphics[width=.48\linewidth]{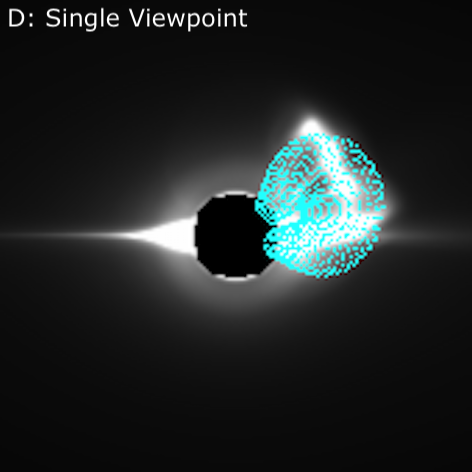}
\includegraphics[width=.48\linewidth]{figures/white_synthetic_placeholder_image.png}\\
\includegraphics[width=.48\linewidth]{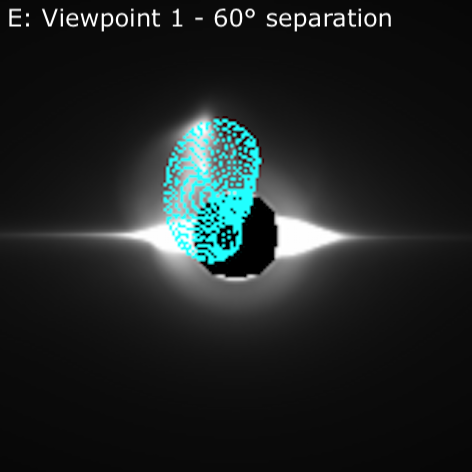}
\includegraphics[width=.48\linewidth]{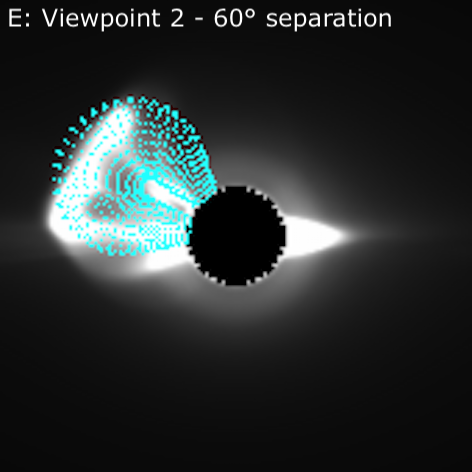}\\
\includegraphics[width=.48\linewidth]{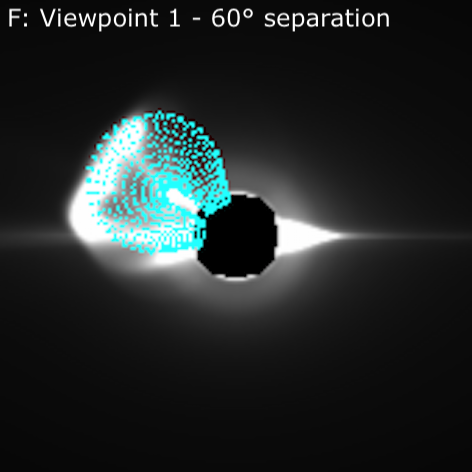}
\includegraphics[width=.48\linewidth]{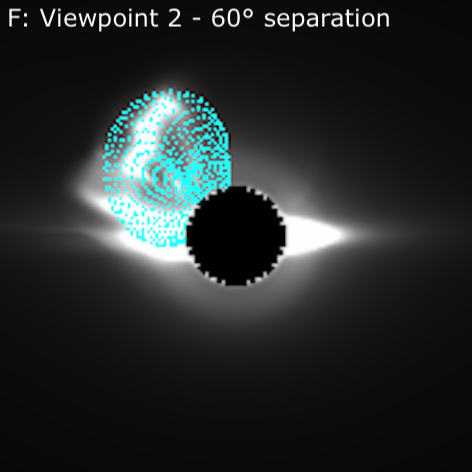}\\
\includegraphics[width=.48\linewidth]{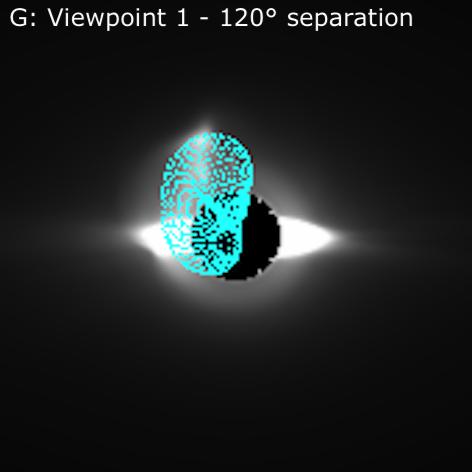}
\includegraphics[width=.48\linewidth]{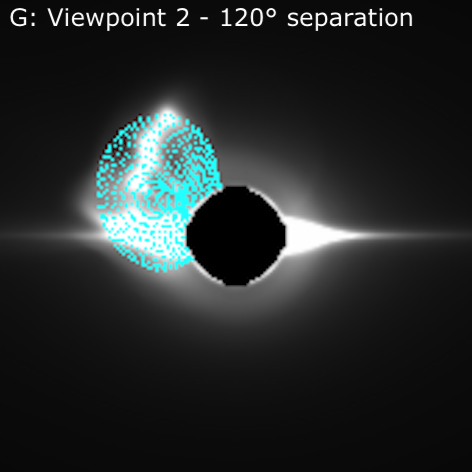}\\
\includegraphics[width=.48\linewidth]{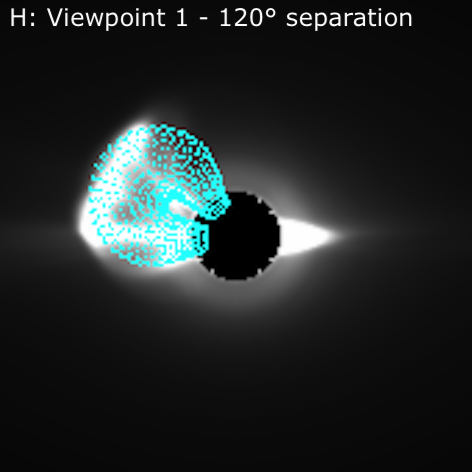}
\includegraphics[width=.48\linewidth]{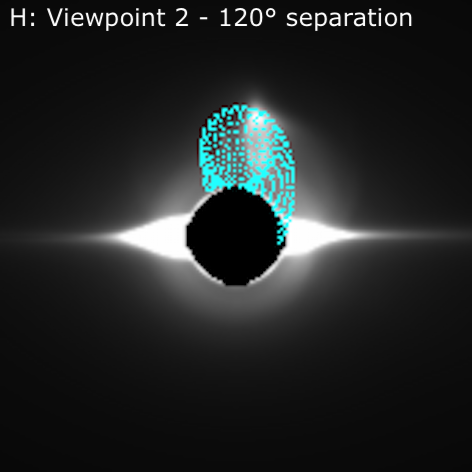}
\caption{Simulated white light data created from an MHD simulation of a hypothetical CME from the different viewpoints shown in Figure \ref{fig:simconfigmhd} and listed in Table \ref{tab:sim}. The images were created for a COR2-like field of view, out to approximately $15\,R_{\odot}$.  Here we have zoomed in to approximately $5\,R_{\odot}$ to better show the CME. The GCS wireframes derived from the MHD truth are also shown on these images. Configurations D, E, F, G and H are shown in rows 1-5.}
\label{fig:syntheticMHD1}
\end{figure}

Figure~\ref{fig:syntheticMHD1} shows the simulated white-light data from the different viewpoints shown in Figure \ref{fig:simconfigmhd} and listed in Table~\ref{tab:sim}. The images were created for a COR2-like field of view, out to approximately $15\,R_{\odot}$. Here we have zoomed into the field of view to approximately $5\,R_{\odot}$ to better show the CME. The GCS wireframes represent the MHD truth listed in Table~\ref{tab:mhdtruth}. Configurations D, E, F, G, and H are shown in rows 1--5. The white-light images show how some other coronal structures could be considered as part of the CME when performing a GCS fitting, resulting in a fit that differs from the ground truth, as discussed in the next section.  We also observe that the CME front is not symmetric and round, as in the GCS croissant shape.

\subsection{CME reconstruction of simulated MHD white-light data}\label{subsec:mhdresults}

Now we turn to the GCS CME reconstruction and related errors of the MHD simulated CME for the five different configurations as presented in Figure~\ref{fig:simconfigmhd}. This is a departure from the synthetic white-light images generated by GCS (see Figure~\ref{fig:syntheticGCS}), for which the GCS geometrical model can be fit perfectly. 
Ten different people reconstructed the simulated CME (ten fits) using the GCS technique for each of the five configurations. As described in the previous section, the ``true" CME parameters were derived from the MHD simulation, where we also determined an uncertainty associated with this derived truth. Still, this is a valuable exercise because, compared to fitting real CME data, the truth is known up to an uncertainty. 

\begin{figure}[ht]
\centering
\includegraphics[width=.98\linewidth]{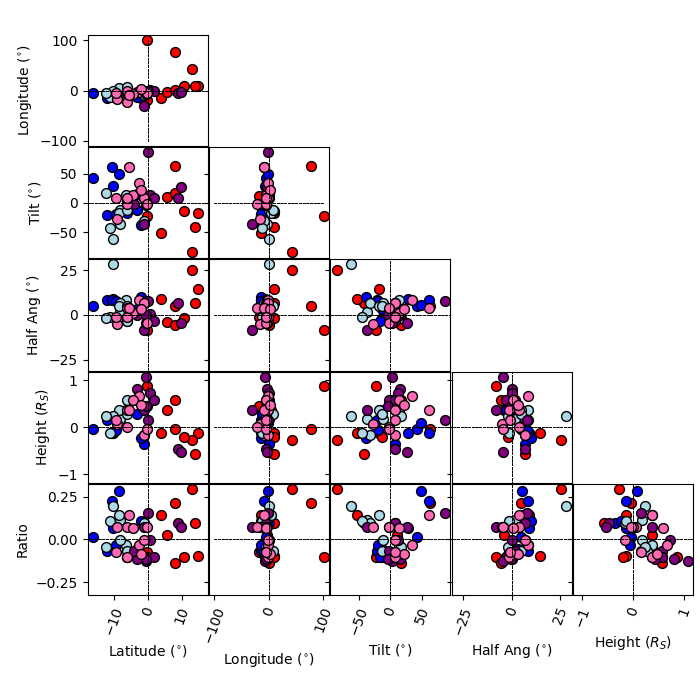} 
\caption{Correlation plots of GCS parameters for each of the 5 configurations (D through H) in Figure \ref{fig:simconfigmhd}. Events D, E, F, G, H correspond to red, blue, light blue, purple and pink, respectively.  Please note the differing axis ranges used for the various parameters.}
\label{fig:triplotsim}
\end{figure}

Figure~\ref{fig:triplotsim} shows the CME fit results for all five configurations (D through H) in the form of a collection of correlation plots between all possible GCS fit parameters (in the same format as Figure~\ref{fig:synt_L5}). All of the values are shown relative to the MHD ground-truth CME parameters, which is the $(0,0)$ point (dashed cross hairs) of each plot. Because each of the five configurations is actually observing the same CME event, this allows us to plot the results together. For this purpose, the measured GCS longitudes (related to the actual white-light header data) were transformed into MHD Carrington longitudes. Each color represents a different configuration (D: red, E: blue, F: light blue, G: purple, H: pink). The single-viewpoint configuration D (red) has the largest offset from the truth values for all parameters, except height, as well as the largest spread for most of the parameters, compared to all of the other configurations. We observe that the spread in reconstructed CME parameters using one viewpoint is much larger than for two viewpoints, particularly for the longitude parameter. Visual examination of the ten GCS reconstructions for configuration D shows that they could be all reasonable fits. These results point toward the critical need of more than one viewpoint to be able to reduce the uncertainty in deriving CME parameters.

\begin{figure*}[ht!]
\centering
\includegraphics[width=.32\linewidth]{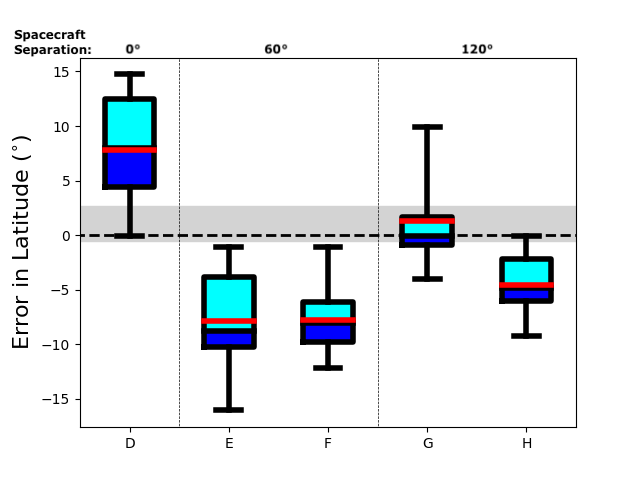}
\includegraphics[width=.32\linewidth]{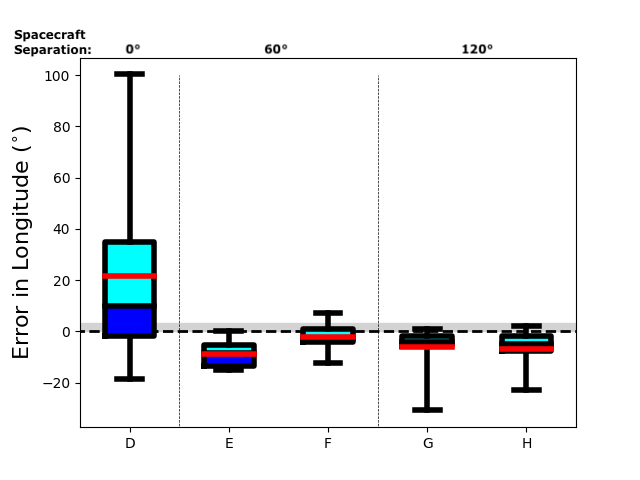}
\includegraphics[width=.32\linewidth]{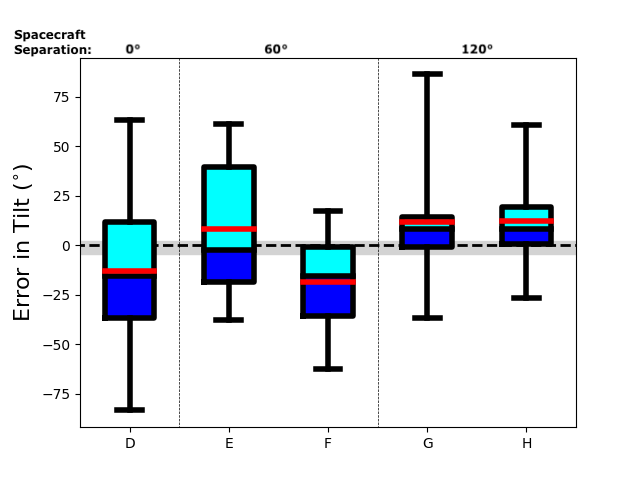}\\
\includegraphics[width=.32\linewidth]{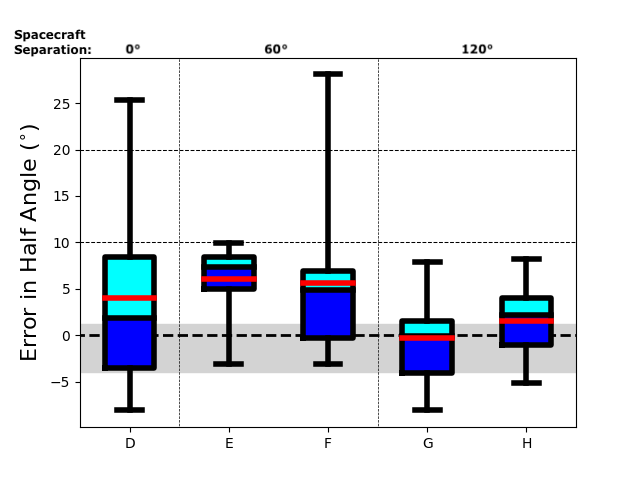}
\includegraphics[width=.32\linewidth]{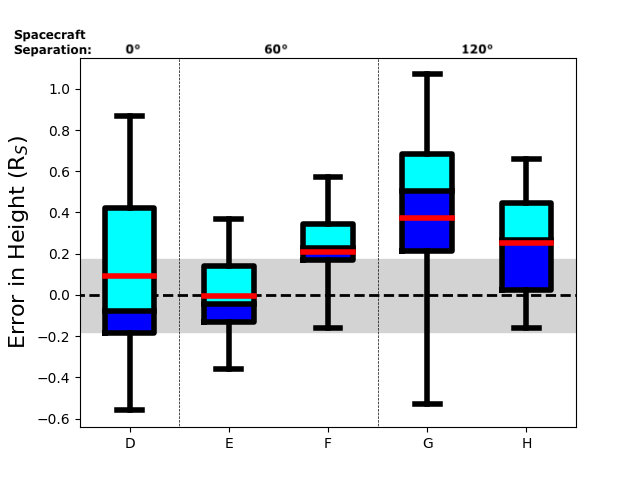}
\includegraphics[width=.32\linewidth]{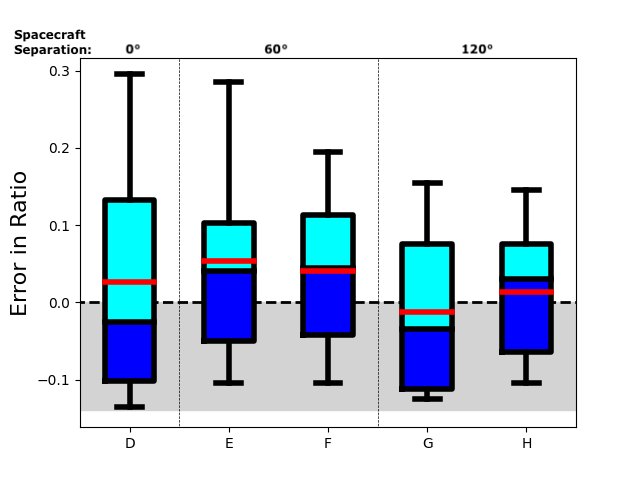}\\
\includegraphics[width=.32\linewidth]{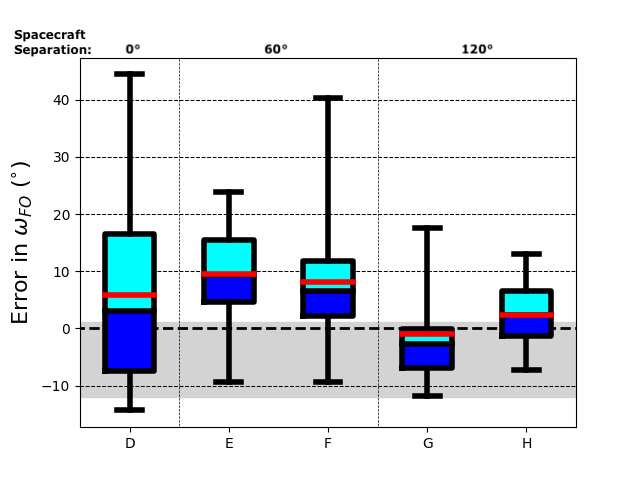}
\includegraphics[width=.32\linewidth]{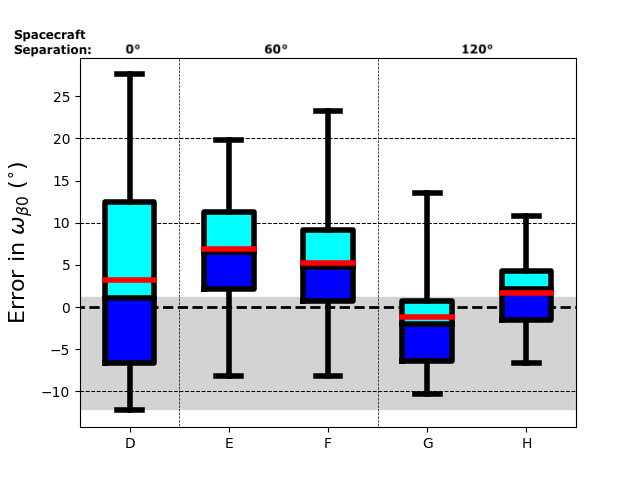}
\includegraphics[width=.32\linewidth]{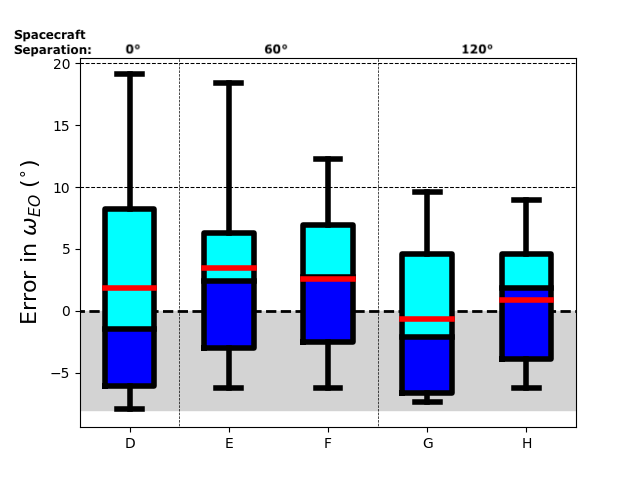}\\
\caption{Box and whisker plots of GCS fit parameters for ten fits relative to the ground truth for the MHD simulated event for configurations D, E, F, G and H (see Figure \ref{fig:simconfigmhd}). The median and mean is shown by the black and red horizontal lines respectively, the box represents the first and third quartiles, and the whiskers show the full range. The grey bars denote the uncertainty of the ground truth as discussed in Section \ref{sec:mhd}. First and second rows: GCS parameters of latitude, longitude, axial tilt, half angle, height of apex, aspect ratio (direct GCS outputs). Third row: face-on width, face-on width at $\beta=0^\circ$, edge-on width (derived parameters).}
\label{fig:boxmhd}
\end{figure*}

Table~\ref{tab:mae_mhd} shows the standard deviations and MAE (relative to the ground truth) for each GCS fit parameter from the simulated MHD white-light images.  In Column 1 we are averaging over all simulated MHD configurations (D -- H), and we also show the minimum and maximum MAE. Column 2 shows the standard deviations and MAE for single viewpoint configuration D. Column 3 shows the values averaged over configurations E and F in which two viewpoints are separated by 60$^{\circ}$ and in Column 4 they are averaged over configurations G and H in which two viewpoints are separated by 120$^{\circ}$.  The trends in this table are also clearly visible in the box and whisker plots shown in Figure \ref{fig:boxmhd}.
The maximum MAE values are nearly all from the single viewpoint configuration D, while the minimum MAE values are from a mix of the two viewpoint configurations E--H.  The mean standard deviations (spread of the fits) and the MAE are largest for the CME tilt and longitude, followed by latitude, half-angle, ratio, and height. While this is from a specific simulation, these results set constrains on the size of the error bars to consider for CME parameters from reconstructions of actual white-light data. For this specific case, we note that the spread for the tilt is extremely high compared to other values. We will come back to this point later on and discuss in more detail why this is the case.

\begin{table*}[th!]
    \centering
    \begin{tabular}{l|r|r|r|r|r|r|r|r|r|r|}
         & \multicolumn{4}{c|}{All Configurations} & \multicolumn{2}{c|}{One Viewpoint} & \multicolumn{4}{c|}{Two Viewpoints}  \\
         Simulated MHD & \multicolumn{4}{c|}{D -- H} & \multicolumn{2}{c|}{D} & \multicolumn{2}{c|}{E, F {\scriptsize 60$^\circ$ separation}}  & \multicolumn{2}{c|}{G, H {\scriptsize 120$^\circ$ separation}}  \\ \hline
         & Average & \multicolumn{3}{c|}{MAE} & \multicolumn{2}{c|}{Average} & \multicolumn{2}{c|}{Average} &  & \\
         & \multicolumn{1}{c|}{$\sigma$} &  \multicolumn{1}{c|}{Avg} & \multicolumn{1}{c|}{Min}  & \multicolumn{1}{c|}{Max} & \multicolumn{1}{c|}{$\sigma$}  & \multicolumn{1}{c|}{MAE} & \multicolumn{1}{c|}{$\sigma$} & \multicolumn{1}{c|}{MAE}  & \multicolumn{1}{c|}{$\sigma$} & \multicolumn{1}{c|}{MAE}\\
         \hline\hline
Latitude [$^\circ$] 	&	4.02	&	6.22	&	3.08	&	7.85	&	5.15	&	7.83	&	3.76	&	7.81	&	3.71	&	3.84	\\
Longitude [$^\circ$] 	&	12.81	&	11.16	&	4.81	&	28.76	&	37.65	&	28.76	&	5.34	&	6.77	&	7.88	&	6.74	\\
Tilt [$^\circ$] 	&	29.64	&	25.39	&	18.39	&	33.2	&	39.07	&	33.2	&	28.72	&	27.5	&	25.83	&	19.37	\\
Half angle ($\alpha$) [$^\circ$] 	&	6.17	&	5.86	&	3.75	&	8.11	&	9.79	&	8.11	&	6	&	6.75	&	4.54	&	3.84	\\
Height [$R_{\odot}$] 	&	0.32	&	0.33	&	0.18	&	0.57	&	0.42	&	0.36	&	0.21	&	0.22	&	0.38	&	0.43	\\
Ratio ($\kappa$) 	&	0.109	&	0.098	&	0.074	&	0.129	&	0.146	&	0.129	&	0.11	&	0.1	&	0.09	&	0.08	\\
$\omega_{\mathrm{EO}}/2 = \delta = \arcsin\kappa$ 	&	6.26	&	5.62	&	4.24	&	7.41	&	8.4	&	7.41	&	6.32	&	5.74	&	5.16	&	4.59	\\
$\omega_{\mathrm{FO}}/2 =\alpha + \delta$ 	&	12.43	&	11.48	&	7.99	&	15.52	&	18.19	&	15.52	&	12.32	&	12.49	&	9.7	&	8.43	\\
$\omega_{\mathrm{FO}(\beta=0^\circ)}/2$ 	&	6.18	&	5.71	&	3.99	&	7.68	&	8.96	&	7.68	&	6.12	&	6.2	&	4.84	&	4.2	\\
    \end{tabular}
    \caption{Mean absolute errors (MAE) and standard deviations ($\sigma$) for each GCS fit parameter from the simulated MHD white-light images.  Column 1: averaged over all simulated MHD configurations (D -- H), and we also show the minimum and maximum MAE. Column 2: standard deviations and MAE the single viewpoint configuration D. Column 3: standard deviations and MAE averaged over configurations E and F in which two viewpoints are separated by 60$^{\circ}$. Column 4: standard deviations and MAE averaged over configurations G and H in which two viewpoints are separated by 120$^{\circ}$.  The trends in this table are also clearly visible in the box and whisker plots shown in Figure \ref{fig:boxmhd}.}
\label{tab:mae_mhd}
\end{table*}

The fitting errors for each configuration are further explored in Figure~\ref{fig:boxmhd} in the form of box and whisker plots of the CME parameters. The format of the box and whisker plots is similar to Section~\ref{subsec:synt_gcs}; however, here we have added grey bars to denote the uncertainty of the ground truth as discussed in Section \ref{sec:mhd}. The configurations are labeled on the x-axis: single viewpoint - D, 60$^\circ$ separation - E and F, 120$^\circ$ separation - G and H. 

We notice from Figures~\ref{fig:triplotsim} and \ref{fig:boxmhd} that a few parameters have a larger error than others; the most eye-catching are the longitude and the tilt. We have further investigated the data to determine possible reasons for these results. First, the team reviewed all of the individual fits and deemed them all reasonable and equally valid. Next, we considered other possible reasons due to each observer fitting different CME features or different fitting approaches. For this purpose, we have visually compared the GCS fits for observers that underestimated the longitude with those that estimated the longitude closest to the ground truth value for each of the five configurations. The leading edge of the simulated white-light CME is not symmetric and round like the GCS croissant shape, and the southern edge of the CME is at a larger height compared to the CME nose, creating a flattened front (see Figure \ref{fig:simfigure}). Upon inspection, we observed that observers who underestimated the longitude the most tended to fit the height of the GCS model to the CME edges, which is above the actual CME leading edge apex. On the other hand, the other observers matched the GCS apex close to the actual CME leading edge to encompass more of the leading edge into the croissant and the southern edge is above the GCS shape (this is clearly visible in Figure~\ref{fig:simfigure}). We also found some fits where the observer was fitting background solar wind features as part of the CME, which also led to larger errors in longitude.  This specific issue could have been mitigated had we provided the observers with the simulated white-light data time series instead of just a single time step. Finally, looking at the tilt, we found one observer who consistently fit a low-tilt croissant and there were a few more scattered low-tilt fits by other observers. Since in the low-tilt-fit one is looking at the croissant predominantly edge-on, these observers consistently compensated by using a higher ratio ($\kappa$) to fit the leading edge of the CME. If we do not consider these low-tilt/high-ratio fits, the spread in both tilt and ratio decreases. Overall, we conclude that the spread between observers is generally explained by them fitting different CME features or taking a different approach to the fitting.


\section{Discussion} \label{sec:discussion}

CME reconstructions from both the single-viewpoint synthetic GCS and simulated MHD white-light data had largest errors and spreads overall. In both cases, as the number of viewpoints increased, the errors decreased.  Under both scenarios of synthetic white-light data, as the number of viewpoints increased from one to two, the errors decreased by approximately 4$^\circ$ in latitude, 22$^\circ$ in longitude, 14$^\circ$ in tilt, 10$^\circ$ in half-angle, 1\,$R_{\odot}$ in height, and 0.05 in the ratio $\kappa$. We did not find a significant decrease in errors when going from two to three viewpoints for our specific hypothetical three spacecraft scenario using synthetic GCS white-light data. Note that this result is obtained for a set-up with all three viewpoints within the ecliptic plane (where all coronagraph imagery has been available to date). Results may in fact differ if a viewpoint at an out-of-ecliptic location is considered.

Over all configurations and numbers of viewpoints, the standard deviations and MAE in deriving the CME parameters are significantly higher in the case of the simulated MHD white-light data (Table~\ref{tab:mae_mhd}), compared to those from the synthetic white-light images generated by the GCS model (Table~\ref{tab:mae_synth}), except for the half-angle and height. The values in these tables are a starting point for quantifying the error in CME parameters from white-light reconstructions. For example, we can make an estimate by rounding the maximum errors across both tables:
$\Delta\theta$ (latitude)=${6^\circ}_{	-	3^\circ	}^{	+	2^\circ	}$,
$\Delta\phi$ (longitude)=${11^\circ}_{	-	6^\circ	}^{	+	18^\circ	}$,
$\Delta\gamma$ (tilt)=${25^\circ}	_{	-	7^\circ	}^{	+	8^\circ	}$,
$\Delta\alpha$ (half-angle)=${10^\circ}_{	-	6^\circ}^{	+	12^\circ}$,
$\Delta h$ (height)=$0.6	_{	-	0.4	}^{	+	1.2	}$\,$R_{\odot}$,
$\Delta\kappa$ (ratio)=$0.1	_{	-	0.02	}^{	+	0.03	}$, 
$\omega_{\mathrm{EO}}/2=$	5.6${^\circ}_{-	1.38	^\circ}^{+	1.79	^\circ}$,
$\omega_{\mathrm{FO}}/2=$	12.7${^\circ}_{-	4.71	^\circ}^{+	14.61	^\circ}$,
$\omega_{\mathrm{FO}(\beta=0^\circ)}/2=$	6.3${^\circ}_{-	2.27	^\circ}^{+	6.51	^\circ}$.

We can also compare individual observer consistency between synthetic GCS and simulated MHD white-light images.  We ask: Do observers tend to stay within the same range of parameters for ratio ($\kappa$) and half-angle ($\alpha$)? We found that about half of the observers measured similar ratios for both the GCS and the MHD fits that are very close to the default values from the \citet{thernisien2009} analysis (0.4) with a spread of about 0.1. The rest of the observers did not stay within this range and fitted significantly different ratio values for the GCS-generated versus the MHD-generated white-light images. As such, there is no difference between the GCS-generated and MHD-generated fittings, but overall observers tend to stay within the 0.2 to 0.6 range, with some observers staying true to the default value, not deviating too much. This bias towards the default value is likely due to the starting ratio value of GCS fitting tool, but also because in many situations there are not enough observational constraints to justify changing the ratio (for instance, when the CME is on the limb and the tilt angle ranges in intermediate values). In fact, a proper determination of the ratio can only be obtained using simultaneous observations of CMEs away from the ecliptic as performed by \citet{cremades2020}. In a sample of 12 CMEs, they found a median value of 0.33 for the ratio. One might argue, however, that these CMEs might not be representative of the full population since they are slower events associated mostly with quiescent filaments. The analysis of projected speeds of CMEs (for the lateral growth and centroid variation) from \citet{balmaceda2020}, on the other hand, suggests higher ratio values with a median of $\kappa$=0.7 for a sample of 333 CMEs. Further investigations/studies in this direction are needed. In this regard, Solar Orbiter will provide helpful new views away from the ecliptic. For the $\alpha$ half-angle GCS parameter, observers tend to be less conservative and use a wide range of values. This conclusion might be something for space weather forecasting centers to consider when training their forecasters.

We discussed the individual observer synthetic GCS correlation plots color-coded by observer in Section~\ref{subsec:synt_gcs}, Figure~\ref{fig:synt_L5}, and the individual observer simulated MHD fitting biases in Section~\ref{subsec:mhdresults}. In both cases we saw that while each observer on its own creates subjective spread in parameters, there is a similar spread generated between observers that fit different features of the CME. Arguably, an ideal scenario would be for an experienced observer to derive the best CME parameters by fitting the ``correct" CME features that matches the ``true" CME parameters.  However, since neither the ``correct" features or ``true" parameters are known, we conclude that it is important to consider fitting different CME features when generating an ensemble of CME parameters from white-light reconstructions.

We note that another important factor to consider when performing CME reconstructions is the processing of the white-light data. In this study we used direct white-light images, not running- or base-difference images that are commonly used on actual white-light data. We will examine the impacts of image processing on CME reconstructions in a follow-up study, using real CME events and white-light data.

Finally, we note that the height of the CME fits of the synthetic GCS white-light data was around 10\,$R_{\odot}$, compared to around 4\,$R_{\odot}$ for the simulated MHD white-light, and this could have an impact on the uncertainty of the CME fits.  Experienced observers on the team tend to prefer deriving CME parameters at a height of approximately 10 to 15\,$R_{\odot}$.  We examined the MHD simulation for a later time step at a height of about 10 $R_{\odot}$ and found that the MHD truth CME parameters remained largely the same, except for some slight increase in longitude by 10$^\circ$, and slight tilt in the clockwise direction (but still remaining largely vertical). Therefore we conclude that it is still reasonable for us to compare the CME fitting errors between the synthetic GCS and simulated MHD white-light data at different heights.


\section{Conclusions} \label{sec:conclusions}

We discussed what errors are introduced when trying to derive CME kinematics from observations. While in the past several studies have focused on comparing different CME reconstruction methods \citep[e.g.,][]{mierla2010} as well as determining trends in over- and/or underestimation of parameters \citep[e.g.,][]{jang2016}, no published literature to our knowledge has explored the subjectivity of the human-in-the-loop that affects the 3D CME parameters that are obtained from any reconstruction technique. With this work, we took important first steps toward this goal.

Because it is not possible to know the ``true'' geometrical parameters of the detected CME for observational data, we have designed synthetic situations where the ``true'' geometric parameters are known in order to quantify such uncertainties for the first time. We generated synthetic line-of-sight integrated white-light intensity images using the ray-tracing option from the GCS model software itself. We considered three different GCS configurations, each configuration with a different spacecraft separation set up. We performed our analysis on both single and multiple viewpoint scenarios.

From the results, we observe that moving from one viewpoint to two viewpoints, the performed fits tighten around the true values for both synthetic GCS and simulated MHD white-light data, suggesting the critical need for (at least) two viewpoints for coronagraph observations. Under both scenarios of synthetic white-light data, as the number of viewpoints increased from one to two, the errors decreased by approximately 4$^\circ$ in latitude, 22$^\circ$ in longitude, 14$^\circ$ in tilt, 10$^\circ$ in half-angle, 1\,$R_{\odot}$ in height, and 0.05 in the ratio $\kappa$.  Specifically, having only one viewpoint it is very well possible to find visually good fits that have a wide range of parameters (e.g., higher/lower width in combination with lower/higher height). 

As expected, the errors in measured CME parameters are generally significantly higher in the case of the simulated MHD white-light data compared to those from the synthetic white-light images generated by the GCS model. We found the following CME parameter error bars as a starting point for quantifying the minimum error in CME parameters from white-light reconstructions:  
$\Delta\theta$ (latitude)=${6^\circ}_{	-	3^\circ	}^{	+	2^\circ	}$,
$\Delta\phi$ (longitude)=${11^\circ}_{	-	6^\circ	}^{	+	18^\circ	}$,
$\Delta\gamma$ (tilt)=${25^\circ}	_{	-	7^\circ	}^{	+	8^\circ	}$,
$\Delta\alpha$ (half-angle)=${10^\circ}_{	-	6^\circ}^{	+	12^\circ}$,
$\Delta h$ (height)=$0.6	_{	-	0.4	}^{	+	1.2	}$\,$R_{\odot}$,
$\Delta\kappa$ (ratio)=$0.1	_{	-	0.02	}^{	+	0.03	}$, 
$\omega_{\mathrm{EO}}/2=$	5.6${^\circ}_{-	1.38	^\circ}^{+	1.79	^\circ}$,
$\omega_{\mathrm{FO}}/2=$	12.7${^\circ}_{-	4.71	^\circ}^{+	14.61	^\circ}$,
$\omega_{\mathrm{FO}(\beta=0^\circ)}/2=$	6.3${^\circ}_{-	2.27	^\circ}^{+	6.51	^\circ}$.

We note that what may seem like a small error in the CME height measurement (0.6 $R_{\odot}$), directly propagates into a larger error for the derived CME speed. Using two consecutive fits of the synthetic GCS white-light data, we determined errors in the CME linear speed and found a MAE range from 10 to 49~km~s$^{-1}$. This represents a lower bound estimate of the errors that arise from determining the CME speed from coronagraph images. We also investigated the effect of the spread in CME arrival times by simulating each set of fits as an ensemble run with the ANTEATR model. Our analysis shows that for the range of considered CME speeds arrival time errors of up to 4 hours are found (representing a lower bound estimate). The CME propagation model's sensitivity determines which parameter has the most impact on the arrival time. In our case, we notice that the half-angle and ratio determined by the GCS model creates the largest difference in arrival time, due to its impact on the drag experienced by the CME in the model. Comparing our results to the ensemble CME parameter sensitivity study of \citet{kay2020}, they found that an accuracy of $5^\circ-10^\circ$ in CME width is necessary to achieve a CME arrival time error of 5 hours or less (for fast and extreme CMEs).  In our study, especially for one viewpoint observations, we find a MAE=${10^\circ}_{	-	6^\circ}^{	+	12^\circ}$ for the half angle, which implies this would lead to a CME arrival time error greater than 5 hours.
 
When using the GCS model to fit the MHD simulated CME, first, we find that in fact, the GCS model is too simple to describe the simulated CME. We expect that for real observed CMEs, especially complex events, this also holds true. Second, we want to remark that extracting GCS parameters from simulation results typically requires a modification of the results and some assumptions. For the specific model used here, this resulted in the determination of the face-on half-width for $\beta=0^\circ$, which is rather straightforward. However, the GCS model is only equivalent to the classic ice-cream cone model \citep{fisher1984} when $\alpha$ is set to zero and as such if a performed fit has a non-zero $\alpha$, a choice will have to be made on how to modify the parameters and this will add an additional error.
 
Apart from the subjectivity of the user performing the fit, other difficulties that introduce errors when performing a fit on real observations arise. We note here a few, but the list is non-exhaustive. Firstly, each user decides for themselves how to process the data. One can decide to use running differences, i.e.\ subtracting two consecutive images for all fits on the studied event, or base differences, i.e.\ subtracting always the same pre-event image; as opposed to using `direct' images as done in the present study. Also, differences in data processing can produce artifacts in the images that may lead to the false interpretation of a feature as part of the CME, or also may enhance faint features that are almost imperceptible in other type of images. Our exercise with the simulated MHD white-light images provides a clear example on how the CME fit parameters may vary from observer to observer. This study represents a lower limit, because when using real white-light data, the error will be always larger. Secondly, the GCS model may not perfectly fit the observed CME shape. As such, the observer decides what feature of the CME to perform the fit on, e.g.\ by fitting the overall shape of the CME the best, or trying to get the frontal part fitting best. Furthermore, the knowledge of the user on the GCS model and its geometric dependencies highly influences the fit they are able to perform.

While it is rather difficult to disentangle all of the possible influences that generate errors in determining the CME kinematics, there are a few things we can do to better estimate this. In this work, we have made first steps towards this by considering a set up where the ``true'' observational values of the detected CME are known. Furthermore, our ISSI team has a study in progress about the effect of the different processing (running vs. base differences), studied on a real CME event. Lastly, the team performed blind fits on different real CME events, so that each event has been fitted multiple times by different people, for the first time creating a broader view on how strong the differences between performed GCS fits are.  The results of the blind fits will then be used as inputs to a variety of CME arrival time models which will allow us to more comprehensively propagate the errors from CME parameters to CME arrival time and impact.


\section{Acknowledgments}

The authors are pleased to acknowledge the International Space Science Institute (ISSI) for their support of International Team no. 480, ``Understanding Our Capabilities in Observing and Modeling Coronal Mass Ejections'' (\url{https://www.issibern.ch/teams/understandcormasseject/}), from which this work originated.
C.V. acknowledges support from the Research Foundation – Flanders, FWO SB PhD fellowship 11ZZ216N.
C.K. is supported by the National Aeronautics and Space Administration under Grant 80NSSC19K0274 issued through the Heliophysics Guest Investigators Program
E. Palmerio acknowledges support from NASA HTMS grant no. 80NSSC20K1274.
L.A.B. and E. Paouris acknowledge support from the LWS Grant 80NSSC19K0069.
M.D. acknowledges support by the Croatian Science Foundation under the project IP-2020-02-9893 (ICOHOSS).
M.M. thanks the European Space Agency (ESA) and the Belgian Federal Science Policy Office (BELSPO) for the support in the framework of the PRODEX Programme.
C.S. acknowledges support from the Research Foundation-Flanders (FWO, strategic base PhD fellowship No. 1S42817N), and from the NASA Living With a Star Jack Eddy Postdoctoral Fellowship Program, administered by UCAR’s Cooperative Programs for the Advancement of Earth System Science (CPAESS) under award No. NNX16AK22G. 
H.C. is member of the Carrera del Investigador Científico (CONICET) and appreciates support from grant PIP 11220200102710CO (CONICET).




\input{journal_macros.sty}
\bibliographystyle{model5-names}
\biboptions{authoryear}
\bibliography{bibliography}

\newpage
\renewcommand{\thesection}{A}

\end{document}